\documentclass[final,5p,times,twocolumn,authoryear]{elsarticle}

\usepackage{amssymb}
\usepackage{graphicx}
\usepackage{subcaption}
\usepackage{newtxtext,newtxmath} 
\usepackage{microtype}

\journal{Icarus}

\begin{document}

\begin{frontmatter}



\title{Giant planet formation in the solar system}


\author[first]{A. Raorane}
\affiliation[first]{organization={Department of Earth and Climate Science, IISER Pune},
            addressline={Dr. Homi Bhabha Rd.}, 
            city={Pune},
            postcode={411008},
            country={India}}
\author[second,seventh]{R. Brasser}
\affiliation[second]{organization={HUN-REN Research Centre for Astronomy and Earth Sciences; MTA Centre of Excellence},
            addressline={15-17 Konkoly Thege Miklos Rd.}, 
            city={Budapest},
            postcode={1121},
            country={Hungary}}
\affiliation[seventh]{organization={Centre for Planetary Habitability, University of Oslo},
            addressline={Sem Saelands vei 2A}, 
            city={Oslo},
            postcode={0371},
            country={Norway}}
\author[third]{S. Matsumura}         
\affiliation[third]{organization={School of Science and Engineering, University of Dundee},
            city={Dundee},
            postcode={DD1 4HN},
            country={UK}}    
\author[fourth]{T. C. H. Lau}   
\affiliation[fourth]{organization={University Observatory, Faculty of Physics, Ludwig-Maximilians-University Munich},
            addressline={Scheinerstr. 1},
            city={Munich},
            postcode={81679},
            country={Germany}}
\author[fifth]{M. H. Lee}            
\affiliation[fifth]{organization={Department of Earth Sciences and Department of Physics, The University of Hong Kong},
            addressline={Pokfulam Rd.}, 
            city={Hong Kong},
            country={Hong Kong}}
\author[sixth]{A. Bouvier}            
\affiliation[sixth]{organization={Bayerisches Geoinstitut, University of Bayreuth},
            addressline={Universitatsstrasse 30}, 
            city={Bayreuth},
            postcode={95447},
            country={Germany}}

\begin{abstract}
The formation history of Jupiter has been of interest due to its ability to shape the solar system's history. Yet little attention has been paid to the formation and growth of Saturn and the other giant planets. Here we explore through $N$-body simulations the implications of the simplest disc and pebble accretion model with steady-state accretion and an assumed ring structure in the disc at 5 AU on the formation of the giant planets in the solar system. We conducted a statistical survey of different disc parameters and initial conditions of the protoplanetary disc to establish which combination best reproduces the present outer solar system.  We examined the effect of the initial planetesimal disc mass, the number of planetesimals and their size-frequency distribution slope, pebble accretion prescription and sticking efficiency on the likelihood of forming gas giants and their orbital distribution. The results reveal that the accretion sticking efficiency is the most sensitive parameter to control the final masses and number of giant planets. We have been unable to replicate the formation of all three types of giant planets in the solar system in a single simulation. The probability distribution of the final location of the giant planets is approximately constant in $\log r$, suggesting there is a slight preference for formation closer to the Sun but no preference for more massive planets to form closer. The eccentricity distribution has a higher mean for more massive planets indicating that systems with more massive planets are more violent. We compute the average formation time for proto-Jupiter to reach 10 Earth masses to be $\langle t_{\rm c,\,J} \rangle= 1.1 \pm 0.3$ Myr and for proto-Saturn $\langle t_{\rm c,\, S} \rangle= 3.3 \pm 0.4$ Myr, while for the ice giants this increases to $\langle t_{\rm c,\, I} \rangle= 4.9 \pm 0.1$~Myr. The formation timescales of the cores of the gas giants are distinct at $>95\%$ confidence, suggesting that they formed sequentially.
\end{abstract}




\end{frontmatter}


\section{Introduction}
\label{introduction}
One of the remarkable features of the solar system is its diverse ensemble of planets. Particularly notable is the presence of giant planets of differing sizes and orbital distance. Planetary formation within protoplanetary discs is a complex process that unfolds across varying temporal and spatial scales \citep[e.g.][]{Morbidelli2012}. An important step of this process is the growth of dust grains, which gradually coalesce into larger meter-sized boulders through successive collisions \citep[e.g.][]{Johansen2007J}.\\

Various mechanisms have been proposed to elucidate the subsequent formation of planetesimals. These mechanisms include gravitational collapse within exceptionally dense regions of the disc \citep{Goldreich1973}, the formation of compact clusters composed of aerodynamically sorted particles \citep{Cuzzi2008}, gravitational instability coupled with collisional accretion \citep{Weidenschilling1980}, and fragmentation processes within self-gravitating protoplanetary discs \citep{rice2006planetesimal}. \citet{Johansen2007J} conducted simulations demonstrating the efficient gravitational collapse of boulders within locally overdense regions situated in the midplane of the disc. This collapse is facilitated by transient high-pressure regions and streaming instability.\\

Once planetesimals are formed, they may merge to form larger bodies. The classical core accretion model presents the theory of giant planet formation that begins with planetesimal accretion \citep{Pollack1996}. However, it estimates that a 5 Earth mass ($M_\oplus$) planetary core would take up to 50 million years (Myr) to form in the region of Jupiter and Saturn for typical solid surface densities (7.5 g cm$^{-2}$) \citep{Pollack1996}. These timescales are much longer than the inferred lifetime of nebular gas, which is between 0.1 Myr and 10 Myr, determined by observations of young stars similar to the Sun \citep{Mamajek2009}. Additionally, \citet{Levison2010} have shown that in the outer solar system, core accretion simulations lead to the early planetesimal-driven migration of the growing cores, resulting in the formation of many hot super-earths and gas giants. These studies suggest that the cores of gas giants, as in the solar system, cannot be formed by planetesimal accretion.\\ 

The Very Large Array Observations have shown large populations of grains with the sizes ranging from millimetres to tens of centimetres in discs around other stars; these objects are referred to as `pebbles' \citep[e.g.][]{Testi2003,Wilner2005}. Micron-sized dust grains can grow into pebbles via coagulation until effectively limited by radial drift or by fragmentation \citep{mumma1993protostars,Weiden1977}. Accretion of these pebbles \citep{Ormel2010,LJ2012} by planetesimals has progressively gained acknowledgment as a viable model for the formation of gas giant planets due to its efficiency in forming planetary cores. \citet{Hughes2017} conducted hydrodynamic simulations with direct particle integration and gas drag coupling to show the growth of planetesimals through efficient accretion of pebbles. \citet{Johansen2010} conducted hydrodynamical simulations to investigate the process of pebble and rock accretion onto nascent protoplanets in gaseous discs. Meanwhile, \citet{Ormel2010} demonstrated the high efficiency of pebble accretion, primarily attributed to the increased cross-section in the settling regime. The term `pebble accretion' was coined by \citet{LJ2012} to characterize the process wherein small pebbles are amassed by existing planetesimals, implying that this rapid accretion might facilitate the formation of giant planet cores.\\

Several studies have conducted numerical simulations to explore pebble accretion models for giant planet formation employing simple disc models. \citet{Levison2015} studied the formation of gas giants where pebbles are formed over the lifetime of the disc, and showed the formation of gas giants between 5 and 15 astronomical units (au) from the Sun within 10 Myr. \citet{Matsumura2017} adapted the Symplectic Massive Body Algorithm (SyMBA) \citep{Duncan1998} to incorporate key planet formation processes, including pebble accretion, planet migration, and gas accretion. However, they did not form any gas giants at the end of the 50-Myr simulations due to the rapid migration of planetary cores to the inner edge of the disc, in contrast to \citet{Levison2015} where migration is not considered. \citet{Matsumura2021} built on the results of \citet{Matsumura2017} and were successful in reproducing orbital distribution trends of extra-solar gas giants. When the same model was applied to the solar system, rapid migration of giant planet cores into inner solar system was seen \citep{Lau2024}.\\

In this work, the implications of the simplest disc and pebble accretion model are being investigated in more detail, with steady-state accretion onto the star and an assumed ringed structure in the disc at 5 au \citep{BM2020}. No additional ringed features are assumed to exist, because the effect of a ringed structure in the disc on pebble accretion is not yet well understood. It is further assumed that no additional material is added to the disc from the interstellar medium as it evolves. From this setup we predominantly form planets whose mass is in between that of Saturn and Jupiter. In this study we have performed many preliminary simulations to constrain the input parameters. We only report the results for production simulations with a limited range of input parameters and computational methods and approximations. More detailed studies exploring phase space and different methods are reserved for future work.\\

The paper is structured in the following manner. In Section \ref{theory}, we describe the model used, followed by the numerical methods and initial conditions for the simulations in Section \ref{ic}. In Section \ref{results}, we illustrate the orbital distribution of the giant planets formed within our systems. Additionally, we depict the impact of varying parameter values related to the planetesimal disc and accretion efficiencies on planet formation probabilities. Finally, we present the timelines for core formation in the cases of Saturn and Jupiter, as well as the duration for the cores to accumulate a gas envelope. We discuss the results further in Section \ref{discuss}, and summarise our findings in Section \ref{summary}.

\section{Theory} \label{theory}
In this work, we used the same disc model as in \cite{Matsumura2021}, which is adopted from \cite{Ida2016} as summarized below.

\subsection{Gas disc}
The disc evolution is described by the diffusion equation \citep{lynden1974evolution} for the disc’s surface mass density $\Sigma_g$
\begin{equation}
    \frac{\partial \Sigma_g}{\partial t} = \frac{1}{r} \frac{\partial}{\partial r} \left[3r^{1/2} \frac{\partial}{\partial r} (\Sigma_g \nu r^{1/2}) \right]
\label{eq:surf_dens}
\end{equation}
where $\nu$ is the disc’s viscosity representing the accretion rate. The gas accretion rate onto the star \citep{SS1973}, $\dot{M}_*$ is related to the gas surface density, $\Sigma$, and scale height of the disc, $H$, via
\begin{equation}
\dot{M}_*= 3\pi \alpha_{\rm acc} \Sigma H^2 \Omega_{\rm K},
\label{eq:dotmstar}
\end{equation}
where $\Omega_{\rm K}=\sqrt{GM_*/r^3}$ is the Kepler frequency. Classically, the parameter $\alpha_{\rm acc}$ is a measure for the global angular momentum transfer of the disc, which is parameterized by \citep{SS1973}
\begin{equation}
\nu=\alpha_{\rm acc} c_s^2\Omega_{\rm K}^{-1}.
\end{equation}
With the assumption that $\nu$ is a power law of $r$ the equation can be solved analytically to yield a self-similar solution, \citep[e.g.][]{Hartmann1998, lynden1974evolution}, which we use here rather than numerically solving equation \ref{eq:surf_dens}. We do not integrate it directly. Following \citet{Ida2018} the work by \citet{Matsumura2021} adopted a two-$\alpha$ disk model which mimics the wind-driven accretion disc with a low-level of turbulence \citep{Bitsch2019}. As pointed out by \citet{Matsumura2021}, $\alpha_{\rm acc}$ is due to the effect that is driving the accretion (be it either magnetic disc winds, viscosity, or something else). In contrast, the effect of disc turbulence is represented by $\alpha_{\rm turb}$, which is generally a small parameter since disc turbulence is expected to be weak \citep{Bai2017}. \\

The disc scale height $H=c_s/\Omega_{\rm K}$, where $c_{\rm s}=\sqrt{\gamma k_BT/2.3 m_p}$ is the sound speed, $k_B=1.381\times10^{-23}$~m$^{2}$~kg~s$^{-1}$~K$^{-2}$ is the Boltzmann constant, $m_p$ is the proton mass, the mean atomic mass of the gas is assumed to be $2.3$ and $\gamma=7/5$ is the ratio of specific heats of the gas molecules. The stellar accretion rate in the self-similar solution decreases with time as \citep{Hartmann1998}
\begin{equation}
 \dot{M}_* = \frac{M_d}{(2q+1)t_{\rm diff}}\left(\frac{t}{t_{\rm diff}}+1\right)^{-(2q+2)/(2q+1)}
\label{eq:mdot}
 \end{equation}
where $M_d$ is the mass of the disc, $t_{\rm diff}$ is the diffusion time, and $q=-{\rm d}\ln T/{\rm d}\ln r$ is the negative temperature gradient. The above equation serves as an inner boundary condition while solving equation \ref{eq:surf_dens} to determine surface density distribution over time. The self-similar equation allows for a self-consistent model where the surface density evolution in the disk is accurately tracked along with the mass accretion onto the star. In the irradiative regime of the disc in the giant planet region $q=3/7$ \citep{Ida2016} (see below) and $(2q+2)/(2q+1)=20/13 \approx 1.53$. Most observed discs have a mass that's about 3\%-8\% of the stellar mass \citep{Manara2016}. For a disc mass of $M_d = 0.05$~$M_\odot$ we have initially $\dot{M}_{*}=5.3\times 10^{-8}$~$M_\odot$~yr$^{-1}$ if $t_{\rm diff}=0.5$~Myr and the accretion rate onto the star becomes approximately $2.5 \times 10^{-9}$~$M_\odot$~yr$^{-1}$ after about 5 Myr of evolution, a little later than the inferred disappearance of the disc from meteorite magnetic fields \citep{Wang2017}. \\

In the giant planet region, the disc mid-plane temperature can be approximated by \citep{Ida2016}
\begin{equation}
T = 150 \Bigl(\frac{L_*}{L_\odot}\Bigr)^{2/7}\Bigl(\frac{M_*}{M_\odot}\Bigr)^{-1/7}\left(\frac{r}{1\,{\rm 
au}}\right)^{-3/7}\; {\rm K},
\label{eq:T_visirr}
\end{equation}
where $r$ is the distance to the Sun. To simplify the formulae, we define
\begin{eqnarray}
\alpha_3 &\equiv& \frac{\alpha_{\rm acc}}{10^{-3}},\\
\dot{M}_{*8} &\equiv& \frac{\dot{M}_*}{10^{-8}\,M_\odot\,{\rm yr}^{-1}}.
\label{eq:definealpha}
\end{eqnarray}
With this temperature profile, which is constant in time -- see equation (\ref{eq:T_visirr}) -- one may compute the reduced scale height $h=H/r$ as
\begin{equation}
h = 0.029 \Bigl(\frac{L_*}{L_\odot}\Bigr)^{1/7}\Bigl(\frac{M_*}{M_\odot}\Bigr)^{-4/7}\Bigl(\frac{r}{1\,{\rm 
au}}\Bigr)^{2/7}.
\label{eq:h_visirr}
\end{equation}
Hence, the gas surface density profile is 
\begin{equation}
\Sigma=1785 \Bigl(\frac{L_*}{L_\odot}\Bigr)^{-2/7}\Bigl(\frac{M_*}{M_\odot}\Bigr)^{9/14}\alpha_3^{-1} 
\dot{M}_{*8}\Bigl(\frac{r}{1\,{\rm au}}\Bigr)^{-15/14}\,{\rm g}\,{\rm cm}^{-2}.
\label{eq:Sigma_visirr}
\end{equation}

Equation~\ref{eq:Sigma_visirr} serves as the initial value for our simulations given the other input parameters. At $t = 0$, this profile can be used to solve equation (\ref{eq:surf_dens}), which decreases with time since $ \dot{M}_{*}$ decreases with time according to equation \ref{eq:mdot}, ensuring consistency in the model. Please note that this ignores the exponential decay of the surface density at large radii in the self-similar solution.\\

\subsection{Planet migration}
In the simulations, the equation of motion incorporates planet-disc interactions based on gas disc torques and dynamical friction. Due to computational and methodological limitations the planetary back-reaction onto the gas is ignored. We follow the prescription of \cite{Ida2020}. The acceleration due to the gas is given by

\begin{equation}
    \frac{d\vec{v}}{dt} = - \frac{v_k}{2\tau_a}\vec{e}_\theta -\frac{v_r}{\tau_e}\vec{e}_r - \frac{v_\theta - v_k}{\tau_e}\vec{e}_\theta -\frac{v_z}{\tau_i}\vec{e}_z
\end{equation}

where, $\vec{v}$ is the planetary velocity evaluated at the instantaneous orbital radius $r$, with $v_r$, $v_\theta$, and $v_z$ being its polar coordinate components defined with the unit vectors $\vec{e}_r$, $\vec{e}_\theta$, and $\vec{e}_z$, respectively. Lastly, $v_K$ is the Keplerian orbital speed evaluated at the instantaneous orbital radius $r$.\\

The evolution timescales for the semi-major axis, eccentricity, and inclination are expressed as follows:

\begin{eqnarray}
    \tau_a &=& -(1+0.04K)\frac{t_{\rm wave}}{2h^2} \\
    &\times&\left[ \frac{\Gamma_L}{\Gamma_0} \left(1 - \frac{1}{C_M}\frac{\Gamma_L}{\Gamma_0} \sqrt{\hat{e}^2 + \hat{i}^2}\right)^{-1}  + \frac{\Gamma_C}{\Gamma_0} \exp \left(-\frac{\sqrt{\hat{e}^2 + \hat{i}^2}}{e_f}\right) \right]^{-1}, \nonumber
\label{eq:ta}
\end{eqnarray}

\begin{equation}
    \tau_e = 1.282(1+0.04K)t_{\rm wave} \left[ 1 + \frac{1}{15} (\hat{e}^2 + \hat{i}^2)^{3/2} \right],
\label{eq:te}
\end{equation}

\begin{equation}
    \tau_i = 1.838(1+0.04K)t_{\rm wave} \left[ 1 + \frac{2}{43} (\hat{e}^2 + \hat{i}^2)^{3/2} \right],
\label{eq:ti}
\end{equation}

where eccentricities and inclinations scaled with $h$ are defined as $\hat{e} = e/h$ and $\hat{i} = i/h$, respectively, and $e_f = 0.01 + h/2$ \citep{FN2014} and $C_M = 6(2p-q+2)=156/7\approx 22.3$ \citep{Ida2020}, where we assumed a steady state accretion disc so that the surface density slope $p =-{\rm d}\ln \Sigma/{\rm d}\ln r = 3/2-q = 15/14$. The quantity $\Gamma/\Gamma_0$ represents a normalized torque, and the subscripts L and C, respectively, correspond to Lindblad and corotation torque. These were computed using the prescription of \cite{Paardekooper2011}, in which the corotation torque also depends on the amount of saturation of the barotropic and entropic parts. The part of the torques dependent on temperature and surface density slopes $q$ and $p$ are $\Gamma_L/\Gamma_0 = 2.50-0.1p+1.7q \approx 1.81$ and $\Gamma_C/\Gamma_0 = 2.73+1.08p+0.87q \approx 1.95$ \citep{Ida2020}. Lastly, $t_{\rm wave}$ is the characteristic time of wave induction in the disc, which propels the migration, and is given in \citet{Tanaka2002} as
\begin{equation}
    t_{\rm wave} = \Bigl(\frac{M_*}{m_p}\Bigr) \Bigl(\frac{M_*}{\Sigma a^2}\Bigr) h^4 \Omega_{K}^{-1}\,
\end{equation}
where $m_p$ is the planetesimal or planet mass and $M_*$ is the mass of the star/Sun. The factor $K$ is related to gap opening and is given by \citep{Kanagawa2018} 

\begin{equation}
K=\Bigl(\frac{m_p}{M_*}\Bigr)^2h^{-5}\alpha_{\rm turb}^{-1}, 
\end{equation}
where $\alpha_{\rm turb}$ represents the strength of the local disc turbulence, which we set to $\alpha_{\rm turb}=10^{-4}$. Gap opening becomes important when $K = 25$ which occurs when

\begin{equation}
m_p \sim 2.6 \Bigl(\frac{\alpha_{\rm turb}}{10^{-4}}\Bigr)^{1/2}\Bigl(\frac{h}{0.03}\Bigr)^{5/2}\Bigl(\frac{r}{1\,{\rm au}}\Bigr)^{5/8}\,M_\oplus.
\end{equation}

The accretion parameter $\alpha_{\rm acc}$ is computed from the disc model rather than treated as an input parameter \citep{Matsumura2021}. \\

In our simulations the torques exerted by the planet on the disc material, and the subsequent modifications to the disc's structure, are not explicitly considered, and may not be fully possible without some merging of N-body and hydrodynamical simulations. Our focus here is on how the disc's torques influence the planet's motion. A more detailed prescription is beyond the scope of this work.

\subsection{Pebble accretion}
\citet{Ida2019} and \citet{Matsumura2021} proposed a pebble accretion model that accounts for planet-disc interactions and the two-$\alpha$ model. The incorporation of these factors has led to efficient accretion, thereby enabling the formation of gas giant cores through pebble accretion within a time frame of 5 Myr. \\

The pebbles are thought to form in a front that sweeps outwards \citep{LJ2014}. The pebble mass flux describes the pebble mass swept up by the pebble formation front per unit time as it sweeps through the disc \citep{Lambrechts2014}, 

\begin{equation}
    \dot{M}_{F} = 2\pi r_{\rm pff} \Sigma_p \frac{d r_{\rm pff}}{dt} 
\label{mdotf}
\end{equation}
where $\Sigma_p$ is the pebble surface density in the pebble migrating region. The pebble formation time, $t_{\rm pff}$, and $r_{\rm pff} $, the radius of the pebble formation front, are computed as \citep{Ida2016}

\begin{equation}
    r_{\rm pff} = 100 f^{2/3} \Bigl(\frac{t}{0.2\,{\rm Myr}}\Bigr)^{2/3} \Bigl(\frac{Z_0}{10^{-2}}\Bigr)^{2/3} \Bigl(\frac{M_*}{M_\odot}\Bigr)^{1/3}\, {\rm au}
\label{eq:rff}
\end{equation}
\begin{equation}
    t_{\rm pff} = 0.2 f^{-1} \Bigl(\frac{Z_0}{10^{-2}}\Bigr)^{-1} \Bigl(\frac{r_{d,0}}{100\,{\rm au}}\Bigr)^{3/2}  \Bigl(\frac{M_*}{M_\odot}\Bigr)^{-1/2}\,{\rm Myr}
\label{eq:tff}
\end{equation}
where $r_{d,0}$ is the disk's outer radius and $Z_0$ is the initial solid-to-gas ratio in the disc and is typically assumed to be 0.01. For high enough speed collisions, grains rebound or fragment rather than coagulate, which is called the bouncing or fragmentation barrier \citep{BW2000}. We model this inefficient coagulation through the sticking efficiency $f$ of dust to pebbles, which is another input parameter. \\

To compute the flux of pebbles in the disc, we used the derivation by \citet{Matsumura2021}, which yields
\begin{eqnarray}
    \dot{M}_F &=& 0.021 f^{2/3} \Bigl[\frac{\ln ({R_{\rm peb}}/{R_0})}{\ln 10^4}\Bigr]^{-1} \Bigl(\frac{T_2}{150\,{\rm K}}\Bigr)^{-1} L^{-2/7}_{*0} M^{8/7}_{*0} \nonumber \\ 
    &\times&\alpha^{-1}_3 \dot{M}_{*8} \Bigl(\frac{Z_0}{0.01}\Bigr)^2 \Bigl(\frac{r_D}{1\,\rm au}\Bigr)^{-4/7}
    \Bigl(\frac{t}{t_{\rm pff}}\Bigr)^{-8/21}\, M_{\oplus}\, {\rm yr}^{-1},
\label{mdotf_final}
\end{eqnarray}

where $T_2=150$~K is the characteristic disc temperature due to stellar radiation, ${r_D}$ is the characteristic disk radius and $R_0$ and $R_{\rm peb}$ are the initial radii of dust particles and the pebbles when they start to migrate, respectively \citep{Matsumura2021}. Integrating the above equation, the total amount of solid material drifting through the planet-forming region over 5 million years is approximately 300~$M_\oplus$ for nominal disc parameters. This value appears high, but is within the range of that of previous studies \citep{Bitsch2019}. It could be reduced by adding in the disc taper suggested by \citet{Satoetal2016} -- which adds a term $(1+t/t_{\rm pff})^{-3/2}$ to equation (\ref{mdotf_final}) -- but early numerical experiments showed that we observe almost no growth of planets because the pebble flux decreases much more rapidly with time, so we removed it. We will explore adding its effect in future studies.\\

The rate of accretion from pebbles onto a planetesimal-sized or larger body is derived to be \citep{Ida2016}

\begin{equation}
\dot{M}_{\rm core} = \varepsilon\dot{M}_F,
\label{eq:accr_rate}
\end{equation}
where $\dot{M_F}$ is the pebble mass flux. The accretion efficiency $\varepsilon$ is given by \citep{Ida2016}
\begin{equation}
\varepsilon = {\rm min}\Bigl[1,\frac{C_{\rm i}C_{\rm b} {\hat b}^2\sqrt{1+4\mathcal{S}^2}}{4\sqrt{2\pi}{\mathcal S}\hat{h}_p}\Bigl(1 + \frac{3{\hat b}}{2\chi\eta}\Bigr)\Bigr], 
\label{eq:ida}
\end{equation}
where $\dot{M}_F$ is the flux of pebbles through the disc; its temporal evolution depends on the gas accretion rate onto the star and the formation efficiency of pebbles. The other quantities in equation (\ref{eq:ida}) are  given by
\begin{eqnarray}
b&=&2\kappa r_H{\mathcal S}^{1/3}\,{\rm min}\Bigl(\sqrt{\frac{3r_H}{\chi\eta r}}{\mathcal 
S}^{1/6},1\Bigr), \nonumber \\
\chi &=&\frac{\sqrt{1+4\mathcal{S}^2}}{1+\mathcal{S}^2}, \nonumber \\
h_p&=&\Bigl(1+\frac{\mathcal{S}}{\alpha_{\rm turb}}\Bigr)^{-1/2}h \nonumber \\
C_{\rm b} &=& \min\Bigl(\sqrt{\frac{8}{\pi}}\frac{h_p}{b}, 1\Bigr), \nonumber \\
\eta&=& \frac{1}{2}h^2 \Bigl | \frac{d \ln P}{d \ln r} \Bigr | =\frac{39}{28} h^2\nonumber \\
\ln \kappa &=& -\left(\frac{{\mathcal S}}{{\mathcal S}^{*}}\right)^{0.65}, \nonumber \\
\mathcal{S}^* &=&{\rm min}\left (2,4 \eta^{-3} \mu \right), \nonumber \\
C_{\rm i} &=& \frac{1}{2}\frac{{\rm erf}(\frac{z+b}{\sqrt{2}h_p})-{\rm erf}(\frac{z-b}{\sqrt{2}h_p})}{{\rm erf}(\frac{b}{\sqrt{2}h_p})}, \nonumber
\label{eq:array}
\end{eqnarray}

where $\mu=m_p/M_*$, $\alpha_{\rm turb}$ is the disc's turbulent viscosity, $r_H = r(\mu/3)^{1/3}$ is the Hill radius, $P$ is the 
gas pressure and $\mathcal{S}$ is the Stokes number. Quantities with a circumflex are scaled by the distance to the star, and $h_p$ is the scale height of the pebbles as they drift sun-ward through the disc.\\

An alternative prescription for the pebble accretion efficiency is given by \citet{OL2018}, who included combined effects of the planet's gravitational attraction and gas drag. They studied the 3D pebble accretion efficiency by considering the effects of eccentricity, inclination, and disc turbulence. When the planetesimal radius is small compared to the scale height of the pebble stream, the accretion is 3D and the efficiency in the settling regime is given by

\begin{equation}
    \varepsilon = \frac{A_3}{\eta h_{\rm p,eff}} \Bigl(\frac{m_p}{M_*}\Bigr) f^2_{\rm set}
\label{ormel}
\end{equation}

where $A_3=0.39$ is a constant, $h_{\rm p,eff}$ is the effective reduced pebble scale height, which accounts for the reduced interaction
between pebble and planetesimal by either turbulent stirring of
pebbles or planetesimal inclination, and $f_{\rm set}=\exp[-0.5(\Delta v/v_*)^2]$ is the settling fraction. We further have $\Delta v={\rm max}(0.76e,\eta)v_K$, which is the approach speed between the pebble and the planetesimal, and the settling velocity $v_*=(m_p/M_*)^{1/3}\mathcal{S}^{-1/3}v_K$, where $v_K=r\Omega_K$ is the Kepler velocity. When $\Delta v \gg v_*$, the accretion is in the ballistic regime, and is highly inefficient. Furthermore $h_{\rm p,eff}\approx h_p$ when $i_p \ll h_p$, and $h_{\rm p,eff} \approx 1.25 i_p$ when $i_p \gg h_p$. When the planetesimal mass grows large enough such that its capture radius becomes comparable to the pebble scale height, the accretion is 2D and the efficiency becomes

\begin{equation}
    \varepsilon = \frac{A_2}{\eta}\Bigl[\frac{m_p}{M_*}\frac{\Delta v}{v_*}\mathcal{S}^{-1}\Bigr]^{1/2} f_{\rm set},
\label{ormel2}
\end{equation}
where $A_2=0.32$. The crossover between the two regimes in the low-eccentricity regime occurs when the mass of the planetesimal satisfies $m_p/M_* =\eta^3{\mathcal S}$. The quantity $f_{\rm set}$ is a complicated function of the planetesimal inclination and eccentricity, and the disc's turbulence, and we refer to \citet{OL2018} for more details. We have implemented their full prescription in the software used in this work. The accretion efficiency for a planetesimal with diameter 500~km at 5~au with nominal disc parameters and $e=0$ is $\varepsilon = (3-5)\times10^{-5}$.\\

Pebble accretion occurs until the planet reaches the so-called pebble isolation mass \citep{Lambrechts2014}, after which it ceases due to turbulent wakes created in the disc by the planet. We are using the pebble isolation mass prescription of \citet{Ataiee2018}, which yields

\begin{equation}
\frac{m_{\rm iso}}{M_*} = h^3\sqrt{0.01+37.3\alpha_{\rm turb}}\Bigl[ 1+0.2\Bigl(\frac{\sqrt{\alpha_{\rm turb}(4+\mathcal{S}^{-2})}}{h} \Bigr)^{0.7}\Bigr],
\end{equation}
which results in $\sim 4$~$M_\oplus$ at 5~au and $\sim 16$~$M_\oplus$ at 25 au for the disc parameters we have chosen. All bodies interior to the ones that have reached the pebble isolation mass will no longer accrete pebbles.

\subsection{Gas envelope accretion}
The nucleated instability model proposes that the formation of giant planets may occur when a solid protoplanet, commonly referred to as a core, reaches a critical mass and triggers rapid accretion of nebular gas, resulting in the formation of a massive gaseous envelope \citep{Stevenson1982,BP1986}. This topic is discussed at length in \citet{Ikoma2000} and we use the equations from that work here. The gas envelope accretion proceeds as

\begin{equation}
\frac{dM_{\rm env}}{dt} \sim \frac{M_{\rm core}}{\tau_g}.
\end{equation}
The growth time of the gaseous envelope mass, $\tau_g$, depends strongly on the core mass of the planet, moderately on the grain opacity, and weakly on the past core accretion process. It is given by

\begin{equation}
    \tau_g = b \Bigl(\frac{M_\oplus}{M_{\rm core}}\Bigr)^c  \Bigl(\frac{\kappa_{\rm gr}}{1\, {\rm cm}^2\,{\rm g}^{-1}}\Bigr)\: {\rm yr}
\label{eq:gasenv}
\end{equation}
where $M_{\rm core}$ is the core mass and $\kappa_{\rm gr}$ is the grain opacity. Typically $\log b =8-10$ and $c=2-4$ \citep{Ikoma2000}; for this study we set $\log b=8$ and $c=3$. The opacity was set to 1~cm$^{2}$~g$^{-1}$; its value does not really matter due to the uncertainties in the values of $b$ and $c$. The above equation depends on core mass rather than the total mass of the planet because the core mass primarily determines the rate at which gas can be accreted onto the planet. Once the core reaches a critical mass, it can efficiently accrete gas from the surrounding disk, leading to the rapid growth of the gaseous envelope \citep{Stevenson1982,BP1986}. \citet{Ikoma2000}, \citet{Ida2004} and \citet{Matsumura2021} tested for different parameter values in their numerical simulations to show gas giant formation. \\ 

For low-mass embryos, the gas accretion rate is controlled by the rate at which the embryo’s envelope can cool \citep{PY2014,Bitsch2015}. For large bodies gas envelope accretion is restricted by gas flow to the star and gap opening in the disc by the growing planet. Here we use only the large-embryo prescription. Combining the factors from the envelope contraction and the disc, gas accretion proceeds as \citep{Ida2018}

\begin{equation}
    \frac{dM_{\rm env}}{dt} = {\rm min}\Bigl[\frac{M_{\rm core}}{\tau_g},\dot{M}_*,f_{\rm gap}\dot{M}_* \Bigr],
\end{equation}
where $f_{\rm gap}$ is a reduction factor due to gap opening of the growing planet \citep{Ida2018}, and is given by

\begin{equation}
f_{\rm gap} = \frac{0.031}{(1+0.04K)h^4\alpha_{\rm acc}}\Bigl(\frac{m_p}{M_*}\Bigr)^{4/3}.
\end{equation}

\subsection{Parameter choice}
The input disc mass and diffusion time are taken from observed disc masses and meteorite magnetic measurements that infer a dissipation time of the solar system's protoplanetary disc by about 4~Myr \citep{Wang2017}. The gas accretion parameters have a range of values in the literature. \citep[e.g.][]{Manara2016}. The initial mass in planetesimals was somewhat arbitrarily chosen, but we wanted to ensure that we had a thousand to a few thousand planetesimals, and that there were large enough bodies that would grow fast enough to form gas giants before the end of the simulations. We were inspired by the initial conditions of \citet{Levison2015} regarding the initial planetesimal mass. Planetesimal formation simulations imply a somewhat shallow size-frequency distribution slope of -1.8 \citep{Johansen2015}, shallower than what we have used here; instead, our choice was informed by \citet{Levison2015} and the size-frequency distribution due to collisional grinding \citep{Dohnanyi1969}. We also noted during the test simulations that we produced no gas giants when the sticking efficiency was lower than 0.5.
%

\section{Methods and initial conditions} \label{ic}
To study the formation of giant planets and their evolution, we ran N-body simulations with SyMBAp, the parallel version of the software package SyMBA \citep{Duncan1998,LL2023}. This code has been modified to include the forces from the gas disc, mass growth due to pebble accretion, and gas envelope accretion. We emphasise that pebble accretion is modelled as a mass addition process at each time step; no physical pebbles are actually integrated in the code \citep{Matsumura2017,Matsumura2021}.\\

The parameter space required to generate the desired gas disc and initial conditions is extensive. The number of parameters utilized in this study is contingent upon the specific goals of the simulations. In Table~\ref{tab:par} we list the free parameters in the problem. We chose to keep the quantities in the top five rows fixed because they are somewhat constrained by observations. The quantities in the bottom five rows are less well constrained, and therefore have been tested here, and their range of values that we have employed. The choices for disc mass, gas diffusion time and gas inner and outer edge were obtained from observational constraints \citep[e.g.][]{Manara2016}, and from low-$N$ simulations starting with 2-25 Ceres-mass planetary embryos, and tabulating which combinations yielded gas giant planets.\\

\begin{table*}
\centering
\caption{List of parameters and their range of values}
\renewcommand{\arraystretch}{1.5}
\small
\begin{tabular}{|c|p{2.5cm}|c|c|c|c|}
\hline
\textbf{No.} & \textbf{Parameter} & \textbf{Symbol} & \textbf{Unit} & \textbf{Literature Range} & \textbf{Used value(s)} \\ \hline
1 & Initial mass of gas disc                 & $M_{\rm disc,0}$          & $M_{\odot}$ & $0.01-0.1$ & $0.05$  \\ \hline
2 & Gas disc diffusion timescale     & $t_{\rm diff}$        & yr           & $10^{5-7}$ & $5\times10^5$ \\ \hline
3 & Disc inner and outer edge        & $r_{\rm in}, r_{\rm out}$ & au              & $0.01-5$, $100-500$ & $5$, $100$  \\ \hline
4 & Disc turbulence parameter & $\alpha_{\rm turb}$          & -               & $10^{-(3-5)}$ & $10^{-4}$ \\ \hline
5 & Gas accretion coefficients       & $b, c$            & -, -               & $10^{8-10}$, $2-4$ & $10^8$, 3 \\ \hline
6 & Pebble accretion efficiency prescription     & $\rm peb\_flg$        & -               & - & Ida ($0$), Ormel ($1$) \\ \hline
7 & Sticking efficiency              & $f$               & -               & $0.5-1$ & $0.5$, $0.75$, $1$  \\ \hline
8 & Initial mass in planetesimals      & $m_{\rm tot}$            & $M_{\oplus}$ & $0.01-4$  & $0.5$, $1$, $4$  \\ \hline
9 & Slope of planetesimal size distribution      & $\beta$            & - & $2.5-4.5$ & $2.5, 4.5$ \\ 
10 & Initial number of planetesimals      & $N$            & $10^3$ & -  & $1-1.2, 1.5, 2.4, 3.2, 4.5$ \\ \hline
\end{tabular}
\label{tab:par}
\end{table*}

Following \citet{Levison2015} we began the simulations with an initial planetesimal disc. The assumption of a planetesimal disc is a key component of the model but not universally accepted: the spatial distribution \citep[e.g.][]{Drazkowska2016,Carrera2017,Schoonenberg2018,Lenz2019,Lenz2020} and the condition for the formation of planetesimals through the gravitational collapse caused by the streaming instability \citep[e.g.][]{Carrera2015,Yang2017,Li2021,Gerbig2023} remain active topics of research. \\ 

We ran 840 simulations with 56 different parameter combinations, with most simulations initially having 1000 to 1200 self-gravitating planetesimals i.e. we compute the gravitational force between all bodies. In each simulation, the central object was a Sun-like star and the gas disc's inner boundary was fixed at 5 au, which was a crude attempt to comport with a hypothesised pressure maximum there \citep{BM2020} and to prevent giant planets migrating into the inner solar system. The orbital elements (semi-major axis, eccentricity, inclination, the longitude of the node, the argument of perihelion, and mean anomaly) are uniformly generated with their appropriate multiplicative factors and limits. The initial semi-major axes of the planetesimals ranged from 5 to 25~au.  Every planetesimal was randomly assigned an initial eccentricity within the range of $e = [0, 0.001]$, an associated inclination of $i = 0.5e$ radians, and randomized phase angles.\\

The mass-radius relationship for planetesimals and planet cores that we use here is given by \citet{Seager2007}. The following equation is satisfied by cold terrestrial planets $m_p < 20$ $M_{\oplus}$ of all compositions.

\begin{equation}
    \log \Bigl(\frac{R_p}{3.3\,R_\oplus}\Bigr) = k_1 + \frac{1}{3}\log \Bigl(\frac{m_p}{5.5\,M_\oplus}\Bigr) - k_2\Bigl(\frac{m_p}{5.5\,M_\oplus}\Bigr)^{k_3}
\end{equation}
where $m_p$ and $R$ are the mass and radius of the planet, and where $k_1 =-0.2095$, $k_2=0.0804$, and $k_3=-0.3940$. For planets more massive than 5~$M_\oplus$ we used the prescription $R=1.65(m_p/5\,M_\oplus)^{1/2}$~$R_\oplus$ \citep{WM2014}.\\

The maximum mass of a planetesimal in the initial distribution is estimated by the equation \citep{Matsumura2021}

\begin{equation}
    M_{\rm pltms,init} \simeq 13.8 \Bigl( 1 + 
\frac{\mathcal{S}}{\alpha_{\rm turb}} \Bigr)^{-3/2}  L_{*0}^{3/7}M_{*0}^{-5/7}
\Bigl(\frac{r}{1\,{\rm au}}\Bigr)^{6/7}M_\oplus.
\end{equation}
For reasonable values of $\mathcal{S}\sim 0.1$--1, $\alpha_{\rm turb}=10^{-4}$, $r=10$~au and a bulk density of about 1500 kg m$^{-3}$, the maximum radius of planetesimals is about 2000 km. Our maximum adopted input value is usually 1500 km or 2500 km depending on initial disc mass and size-frequency distribution slope. We placed one planetesimal with a lunar mass and a radius of 2000~km near 5 au with the intention that this could become our Jupiter analogue \citep{BM2020}.\\

The slope of the initial size-frequency distribution of the planetesimals is another free parameter. Both a shallow and a steep size distribution were tested. The total mass of the planetesimal disc, and the minimum and maximum radii of the planetesimals were kept as additional input parameters. The radii are generated from a truncated Pareto distribution using a random number as

\begin{equation}
    r_{\rm p} = r_{\rm min}\zeta^{-1/\beta},
\end{equation}
where $\zeta$ is a random number from an uniform distribution between $(0,1)$ and $\beta$ is the size-frequency distribution slope; if $r_{\rm p} > r_{\rm max}$ the procedure is restarted. \\


Simulations were run for 5~Myr with a time step of 0.1~yr. Bodies were removed when they were closer than 1 au or farther than 100 au from the Sun, or underwent a collision to result in a merger. Between 1 au and 5 au from the Sun we did not model the effects of the gas disc -- migration, gas accretion and damping. \\

\begin{figure*}
  \begin{centering}  
  \includegraphics[width=\textwidth,height=0.4\textwidth]{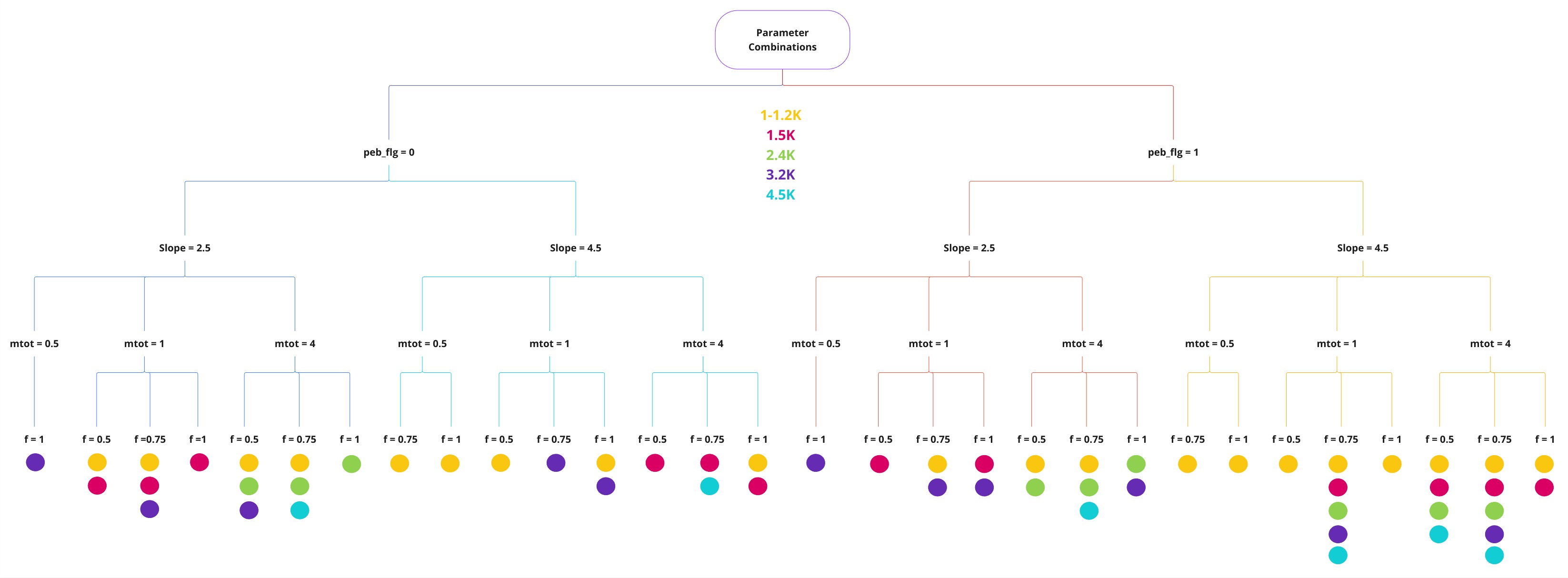}
  \caption['Production Simulation Parameters']{Parameter Combination for Production Simulations. Circles represent a set of 15 runs. The colours correspond to the number of planetesimals.}
  \label{fig:flow}
  \end{centering}
\end{figure*}

\begin{figure}
\resizebox{\hsize}{!}{\includegraphics{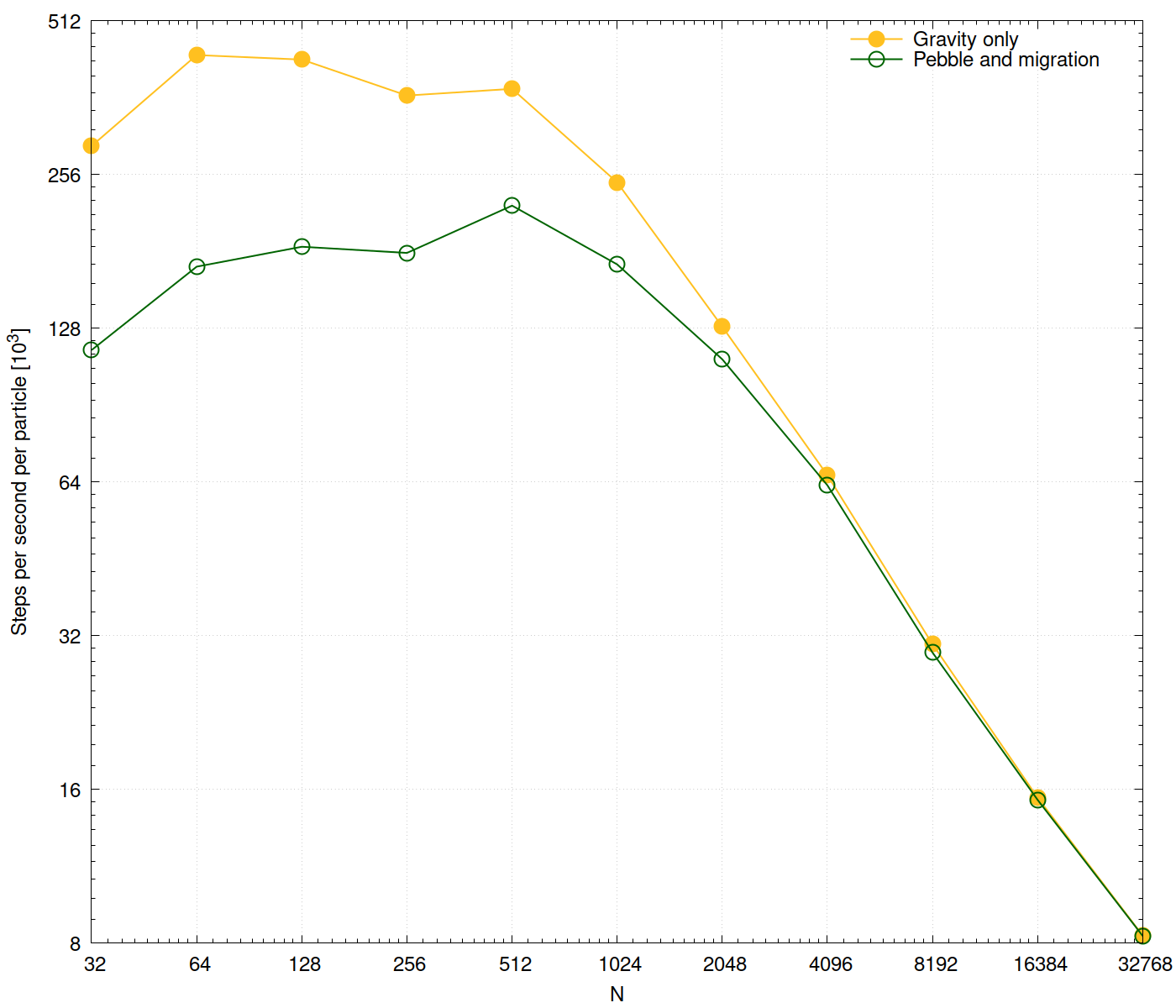}}
\caption['Speed of SymbaP']{Speed of SyMBAp with 8 cores. The solid symbols are for gravity only and the open symbols are for the version used here that includes pebble accretion and planet migration. Simulations were run on AMD EPYC 7H12 CPUs with a clock speed of 3.0 GHz and compiled with GCC 13.2.}
\label{fig:FSGHPC}
\end{figure}

The speed of SyMBAp, using 8 CPU cores on AMD EPYC 7H12 CPUs running at 3~GHz, in both gravity only mode and in pebble accretion with planet migration mode, is shown in Fig.~\ref{fig:FSGHPC}, which depicts the number of steps per particle per second vs the number of bodies. The computational overhead due to pebble accretion and planet migration amounts to about 50\% for $N\leq 512$. For this low number of planetesimals the number of steps per second decreases approximately as $N^{-1}$ while for $N\geq 1024$ it decreases as the expected $N^{-2}$. Due to this rapid decline in speed, for most sets of the simulations, we kept the number of planetesimals between 1000 and 1500, but we have also run some sets with 2400, 3200 and 4500 planetesimals.\\
 
We used 8 cores per simulation. The simulation duration was typically 2 days for 1000 planetesimals and increased as $N^2$, with high-$N$ runs lasting up to 40 days. The parameter combinations used and analyzed in these simulations are presented in Fig.~\ref{fig:flow}; we did not investigate all possible parameter combinations. Fifty-six combinations of the following parameters were analysed: (a) sticking efficiency ($f$), (b) pebble accretion prescription, (c) total initial planetesimal mass in disc ($m_{\rm tot}$), (d) planetesimal size-frequency slope ($\beta$) and (e) initial number of planetesimals $(N)$. 

\section{Results} \label{results}
In this section we present the outcomes of our numerical simulations. Before we get to that, we report on a phenomenon in the simulations that skewed the outcome due to our truncation of the gas disc at 5 au. 

\subsection{Truncated gas accretion}
\begin{figure}
\resizebox{\hsize}{!}{\includegraphics{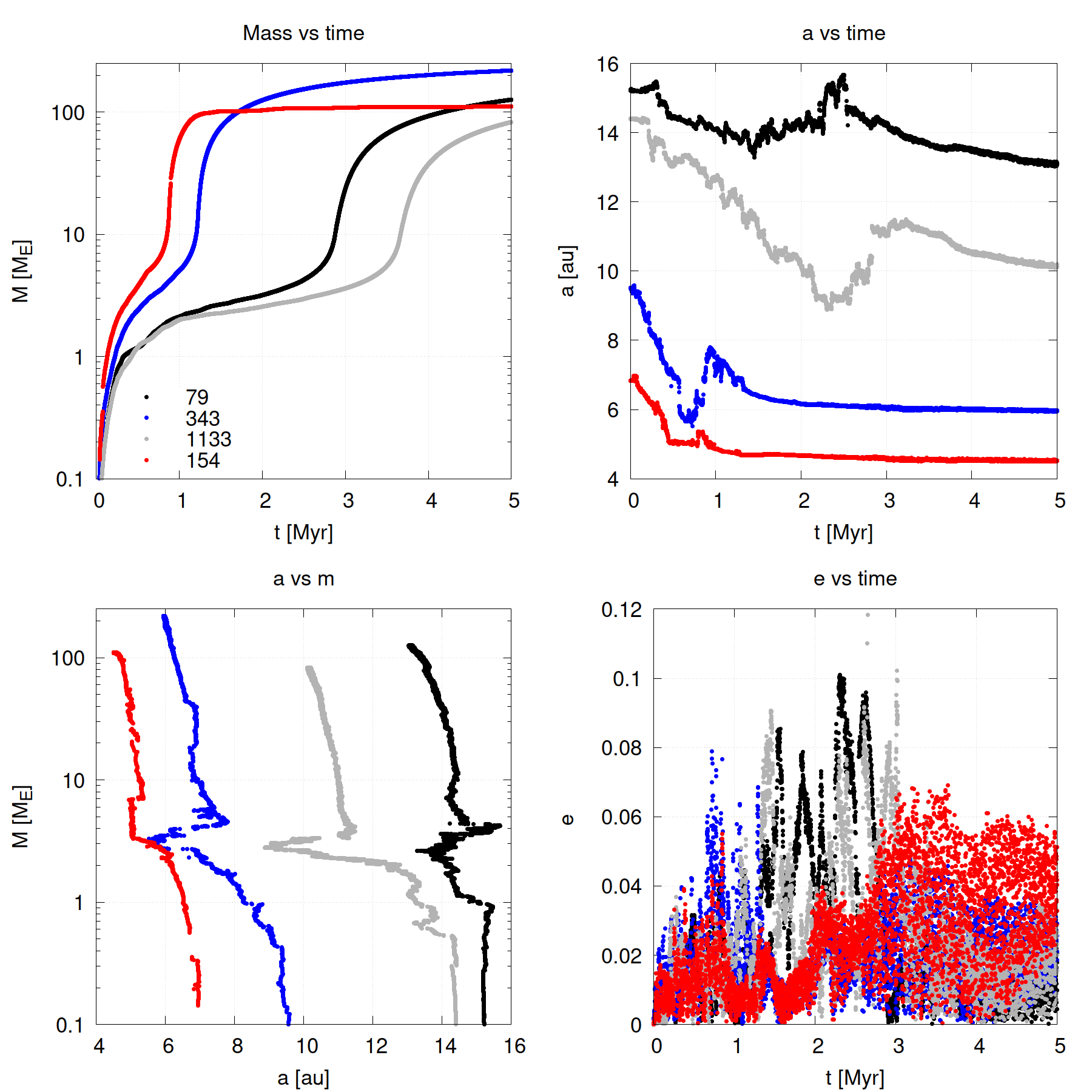}}
\caption['Truncated']{Truncated gas accretion due to the innermost planet being pushed into the cavity. The red (curve) planet is pushed into the cavity, inside 5 au and the gas accretion for it is halted }
\label{fig:trunc}
\end{figure}
We truncated the gas disc in our simulations at 5~au to mimic the pressure maximum postulated by \citet{BM2020}, but all bodies are allowed to venture as close as 1 au to the Sun. At present SyMBAp has no built-in mechanism to compute such a maximum and halt planet migration at the pressure maximum. Thus we opted for this crude approximation that halts planet migration at 5~au with the downside that there is no gas closer to the Sun even though the expectation is that the disc extended closer to the Sun. \\

In our model, we have accounted for planet migration processes. This consideration is crucial because, for a gas giant to be situated at 5 AU, we assume its initial formation occurs farther out in the disk, specifically within the range of 10 to 20 AU. Subsequently, the gas giant migrates inward to its final position at 5 AU. \\

The decision to truncate the disk at 5 AU serves as an artificial barrier designed to prevent the giant planet from migrating further inward toward the Sun. This truncation is not a physical limitation but a methodological choice to constrain the simulation. It is indeed possible within our model for a giant planet to form at 5 AU and cross this truncation radius. Such planets are only removed from the simulation if they approach too close to the Sun, specifically within 1 AU.\\

As we mentioned in Section 3, we place a lunar-mass embryo near 5~au in the hope of triggering the formation of a Jupiter analogue \citep{BM2020}. While the simulations were analysed, we became aware that we produced many Saturn analogues with final semi-major axes $a<5$~au. These planets are situated in a region where no gas is present in the simulation. As a result, this embryo undergoes stunted growth because it settles at the inner edge of the disc, and then another planet that forms later slowly pushes it into the cavity while the planet is still accreting gas. This will eventually halt gas accretion and predominantly produce Saturn analogues inside the cavity. These analogues were not taken into account for the final results. \\

In Fig.~\ref{fig:trunc}, we show that this is indeed the case. This four-panel plot shows the evolution of mass vs. time (top left), semi-major axis vs time (top right), mass vs semi-major axis (bottom left) and eccentricity vs time (bottom right), for a system with four giant planets at the end. Each planet accretes mass at a different rate. The red curve shows the evolution of an embryo that starts near 7 au and then migrates inwards as it grows. After about 1.1 Myr, in the midst of rapidly accreting a gaseous envelope, the gas accretion stops. By this time, the said planet has reached the inner edge of the gas disc because it is pushed into the cavity by the second planet shown in blue; the pushing occurs because they are temporarily trapped in a 2:3 mean motion resonance. Gas accretion for the inner planet is halted while the mass of the outer planet keeps growing. The inner planet's mass is stalled at 110 $M_\oplus$ so that it qualifies as a Saturn analogue (see next subsection). After about 2.5 Myr and 3.5 Myr of evolution, two further planets undergo gas accretion and they end up with masses close to that of Saturn while the planet in blue ends up with a mass in between that of Jupiter and Saturn.
The phenomenon of planets ending up inside the cavity and having their growth artificially halted is rather frequent, and is an artefact of our truncated gas disc at 5 au. Therefore from our results we have removed all giant planets whose final semi-major axis is less than $5$~au.\\

One may argue that if these planets were allowed to accrete gas that they would end up as Jupiter or S-J analogues. This is entirely possible. At this stage we do not have the means to investigate this scenario and we leave that for future work.

\subsection{Orbital distribution of giant planet analogues}
The orbital distribution of giant planets at the end of the simulation shows significant variation among all the trials in terms of the number of giant planets, their orbital elements, the giant planet masses, and the remaining mass of planetesimals.\\ 

\begin{figure}
    \begin{centering}
  \resizebox{!}{0.8\vsize}{\includegraphics{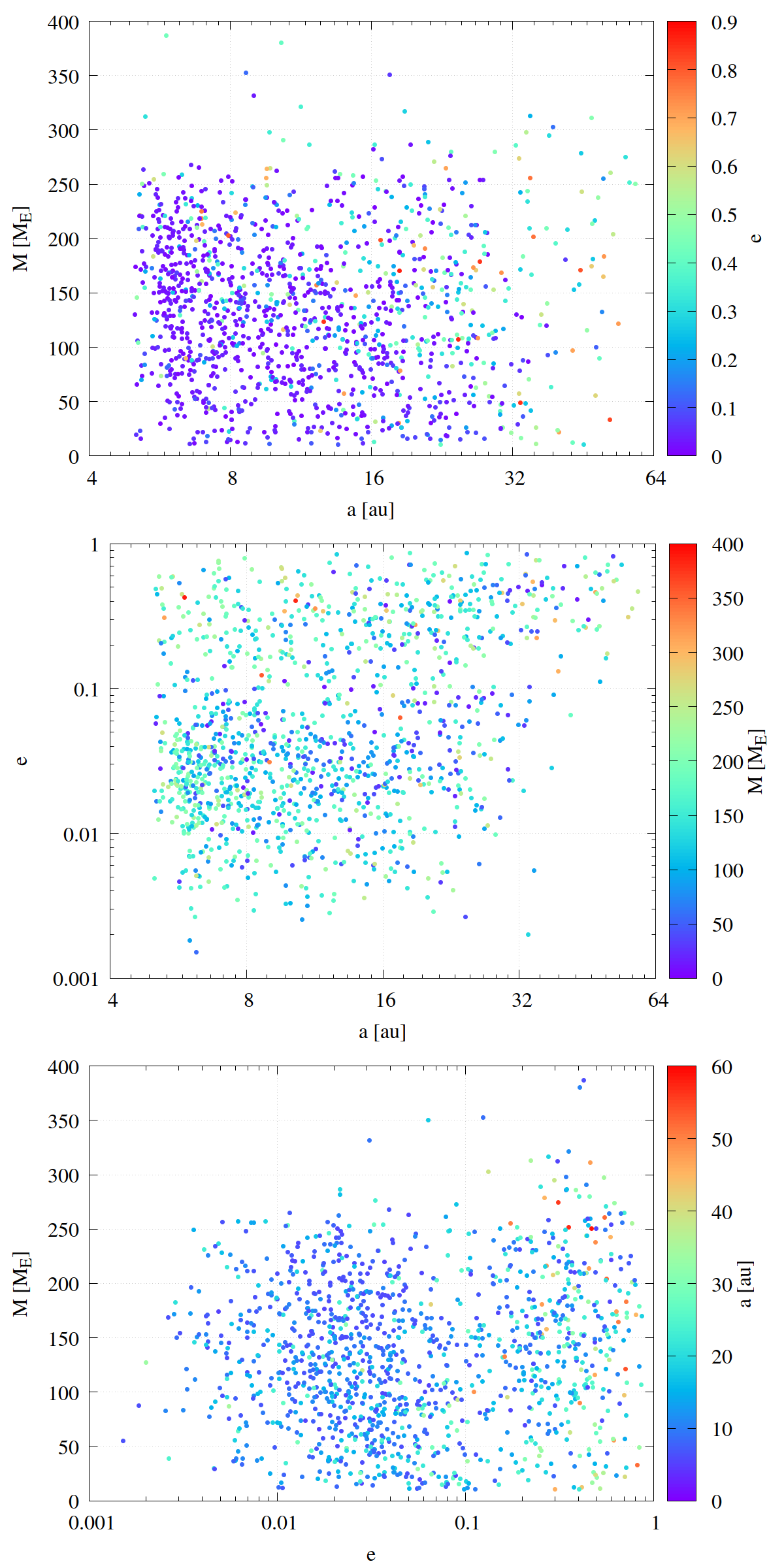}}
   \caption['Orbit Dist']{We show the distribution of giant planet mass, semi-major axis and eccentricity for all planets $m_p > 10 M_\oplus$ over all simulations. The panels are: (a) $m_p-a$ (b) $e-a$ and (c) $m_p-e$. The colouring pertains to the unplotted parameter (see the label next to the colorbar). Any correlation between these parameters is not seen. } 
  \label{fig:orbitdist}
  \end{centering}
\end{figure}

The orbital distribution of planets formed at the end of the simulation is shown in Fig. \ref{fig:orbitdist}, which displays the $m_p-a$, $e-a$, and $m_p-e$ distributions for all planets in all simulations with $m_p > 10 M_\oplus$ and $a>5$~au. We chose a minimum giant planet mass of 10~$M_\oplus$ because this is the approximate pebble isolation mass and from Fig.~\ref{fig:trunc} we observe that gas envelope accretion starts roughly at this mass as well. In total we formed 1467 giant planets with mass $m_p>10$~$M_\oplus$, of which 1158 have a semi-major axis $a>5$~au, so that our outcomes are not constrained by low-number statistics.
There appear to be few planets with a mass $m_p>250$~$M_\oplus$ and, independently, with a semi-major axis $a\gtrsim 30$~au. The latter is not too surprising because we initially placed no planetesimals beyond 25~au. Most planets (66\%) have low eccentricity $e<0.1$. This outcome is expected from systems where planetesimals can damp the eccentricities due to dynamical friction, and where the planets form sufficiently fast that there remains enough gas to assist in the damping. There appears to be no correlation between each of these three parameters. At the end of the simulations there are about 6\% of planets with final eccentricities $e>0.5$ and these are in dynamically unstable systems that have not yet fully evolved. \\

\begin{figure}
\resizebox{\hsize}{!}{\includegraphics{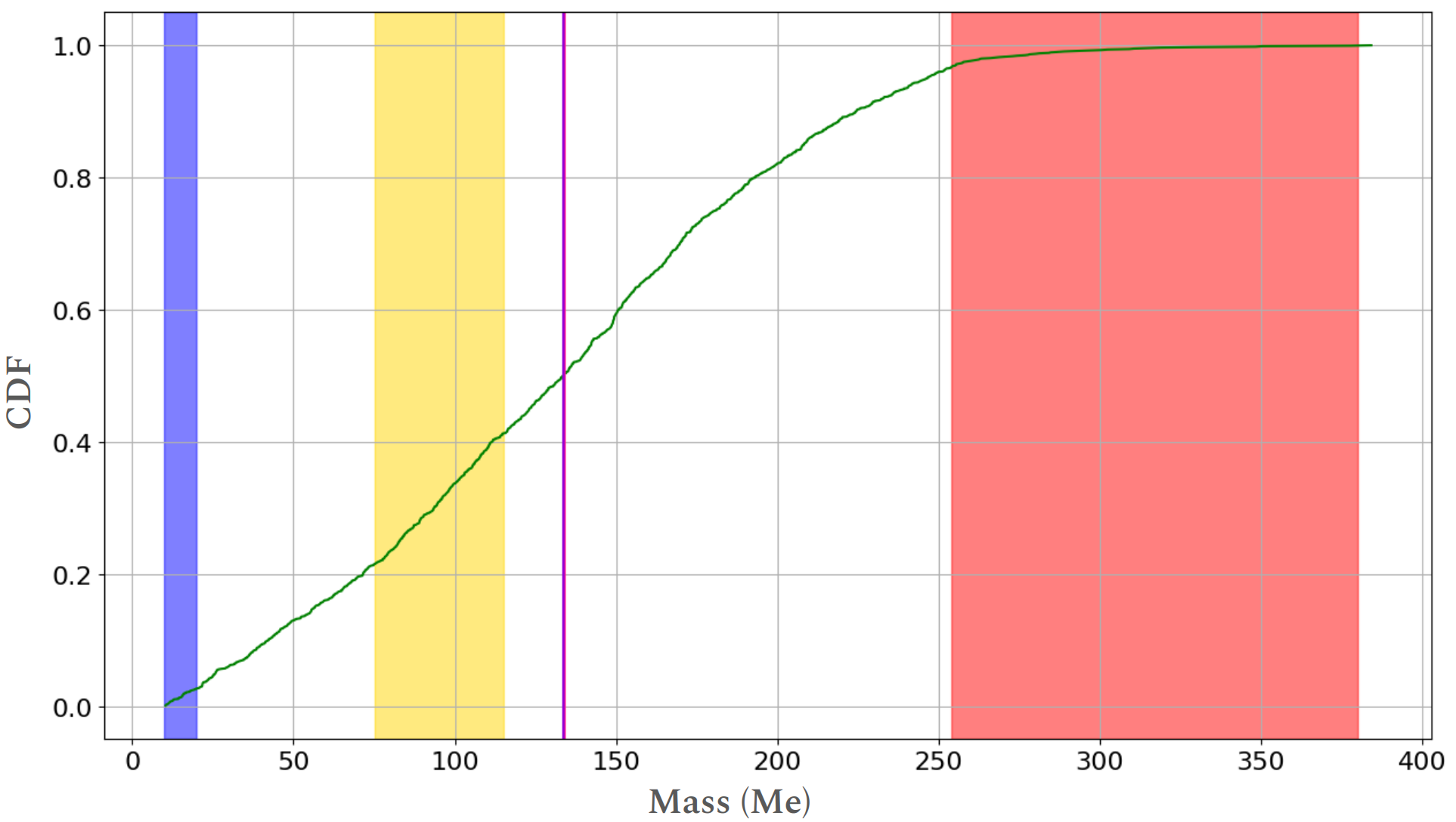}}
\caption['Mass Dist']{Cumulative distribution for the mass in planets with mass more massive than 10$M_\oplus$ and $a >= 5$ au. Both the mean and median mass in planets in each system is around $134\, M_\oplus$. Blue, yellow and red vertical strips indicate range of ice giant, Saturn and Jupiter mass analogues range respectively.}
\label{fig:mcdf}
\end{figure}

In the following, we define the mass range for Saturn analogues to be $75 - 115\, M_\oplus $, $254 - 380\, M_\oplus $ for Jupiter analogues and $10 - 20\, M_\oplus $ for ice giant analogues. As shown in Fig.~\ref{fig:mcdf}, after 5 Myr almost 50 \% of the planet analogues formed have masses between $10 - 135$ $M_\oplus$, while 83 \% have  masses below 200 $M_\oplus$. This implies that very few ice giant and Jupiter analogues are formed in the system, compared to Saturn analogues and planets with masses in between the ice giants and Saturn (I-S), and Saturn and Jupiter (S-J). Overall, 2.6\% of giant planets are ice giant analogues, 17.8\% are I-S analogues, 19.4\% are Saturn analogues, 55.6\% are S-J analogues and 4.4\% are Jupiter analogues. The number of giant planets with $m_p>10$~$M_\oplus$ at the end of the simulations is $\langle N \rangle = 1.7 \pm 1.2$. These percentages are a result of the chosen gas disk parameters, which as mentioned before, were tailored for formation of Saturn. \\

\begin{figure}
\resizebox{\hsize}{!}{\includegraphics{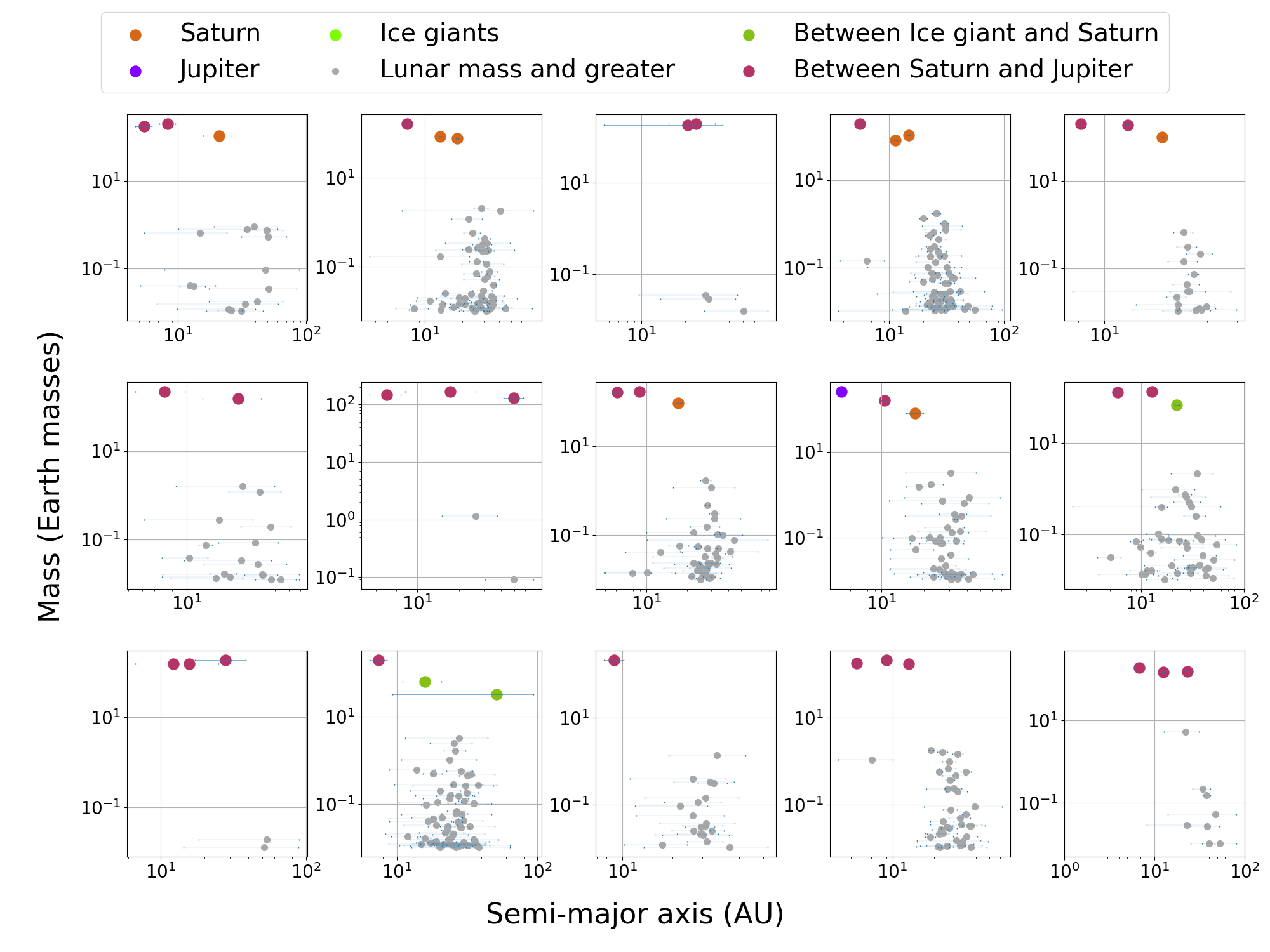}}
\caption['ic16']{Final mass-semimajor axis distribution of a set of simulations with a system yielding both Jupiter and Saturn-mass planets. The middle-leftmost panel has both a Jupiter and a Saturn analogue. The horizontal error bars are a proxy for the eccentricity of the bodies. Parameters: $\rm peb\_flg = 1$, $\beta = 4.5$, $m_{\rm tot} = 4$, $f = 1$ and $N = 1.5K$. For this $f$ value it is difficult to form ice giants (discussed Section 4.3).}
\label{fig:maic16}
\end{figure}

\begin{figure}
\resizebox{\hsize}{!}{\includegraphics{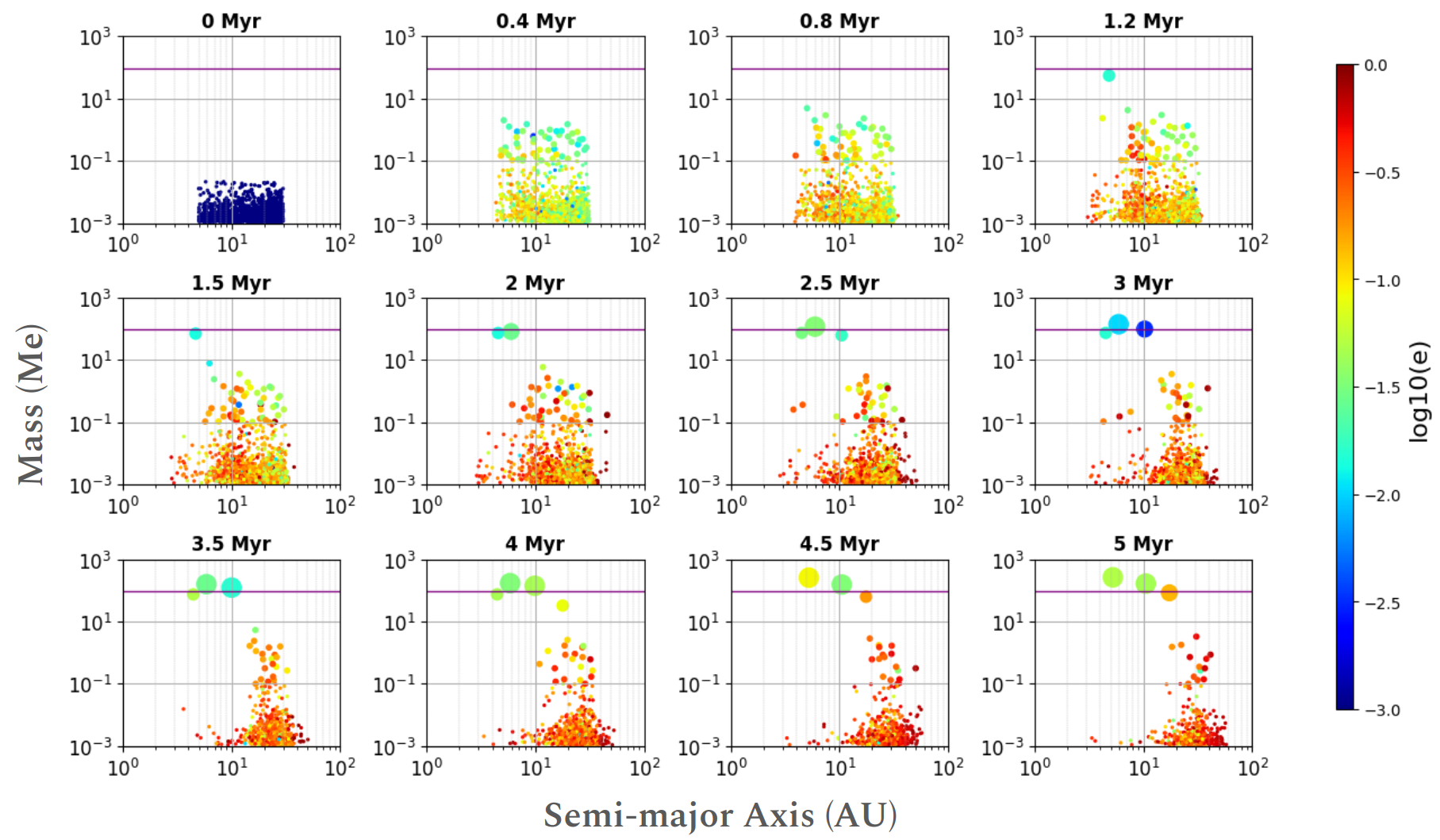}}
\caption['Time Evolution']{Snapshots of the mass and semi-major axis evolution of one simulation producing a Jupiter, Saturn and a SJ analogue. This is the same simulation as the middle leftmost panel in Fig.~\ref{fig:maic16}. The horizontal purple line indicates Saturn mass. The colour coding is a proxy for the eccentricity. High eccentricity values for left-over planetesimals indicate a dynamic system. This is the same simulation as shown in Fig.~\ref{fig:evo}.}
\label{fig:evoexample}
\end{figure}

In all simulations, we have not observed the formation of gas giants and ice giants in the same trial. We demonstrate this with a few figures. In Fig.~\ref{fig:maic16} we show the final mass-semi major axis distribution of one set of simulations wherein we produce at least one Jupiter analogue in the entire set. It is clear that each system is unique, and that the mass in leftover planetesimals depends on the eccentricities of the giant planets at the time the simulations ended. We show the mass-semimajor axis evolution of one simulation producing a Saturn analogue with low eccentricity in Fig.~\ref{fig:evoexample}. The colour coding is proportional to the eccentricity of each body.\\

\begin{figure}
  \resizebox{\hsize}{!}{\includegraphics{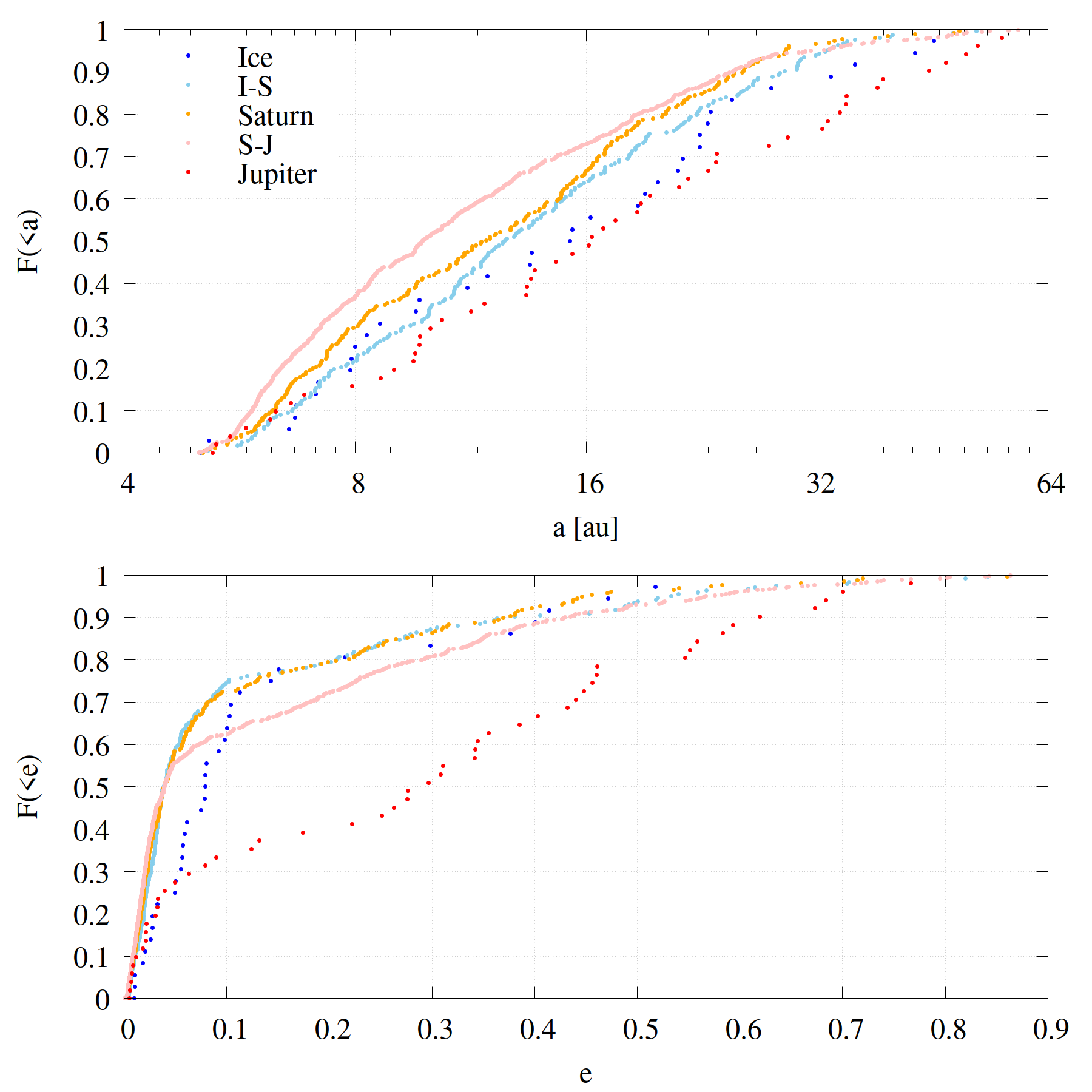}}
   \caption['Orbit CDF']{CDF distribution of (a) a (b) e-a for all planets with $m_p > 10 M_\oplus$ over all simulations. All the planets are approximately evenly placed in $\log a$. Jupiter analogues mostly have higher eccentricities than the other types.} 
  \label{fig:aecdf}
\end{figure}

\begin{figure}[ht]
    \centering
    \begin{subfigure}[b]{\hsize}
        \centering
        \includegraphics[width=\hsize]{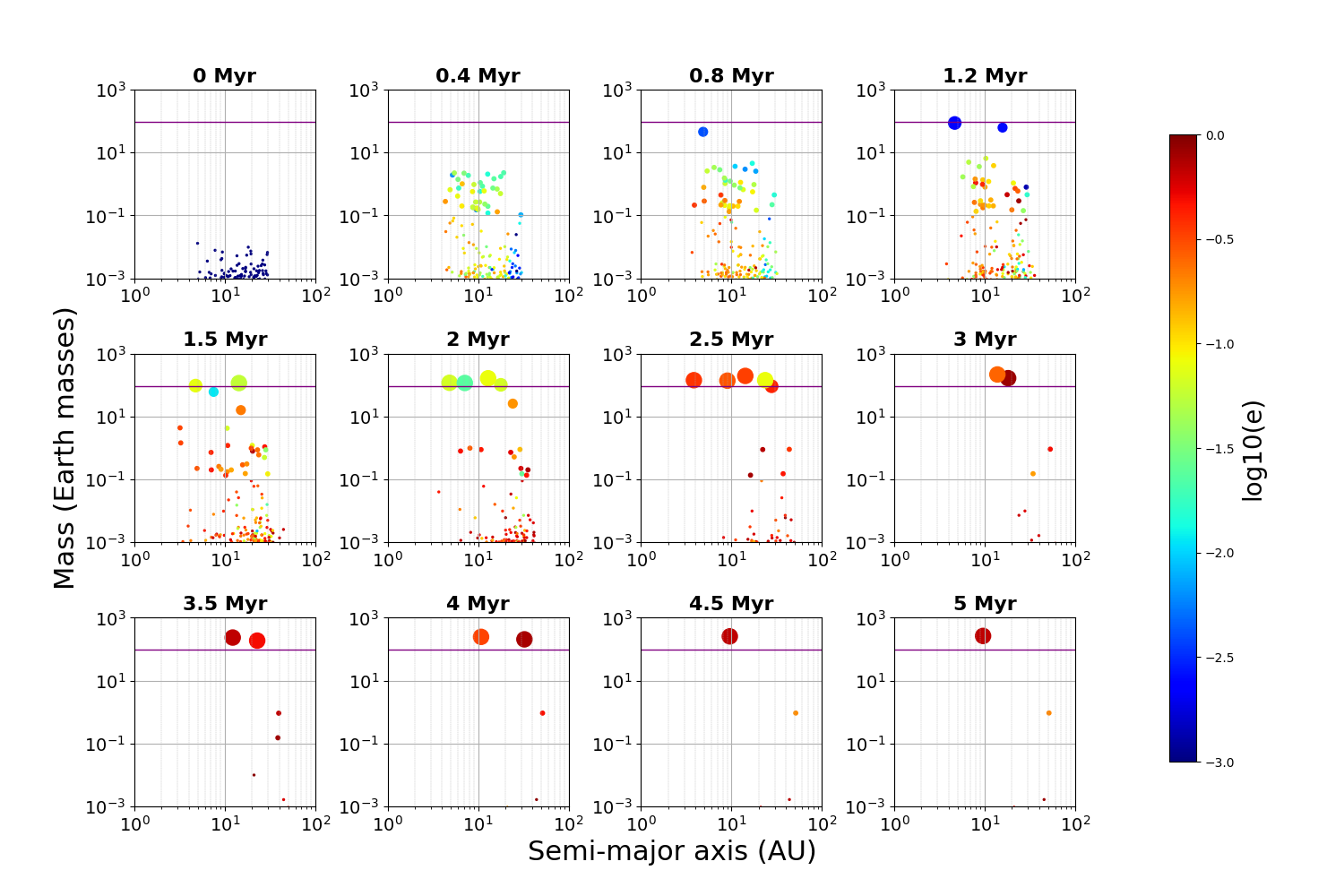}
        \caption{Snapshots of the mass and semi-major axis evolution of one simulation forming an eccentric Jupiter. The horizontal purple line indicates Saturn mass. The colour coding is a proxy for the eccentricity. There are hardly any planetesimals left at the end of 5 Myrs.}
        \label{fig:evojup_a}
    \end{subfigure}

    \vspace{1em} 

    \begin{subfigure}[b]{\hsize}
        \centering
        \includegraphics[width=\hsize]{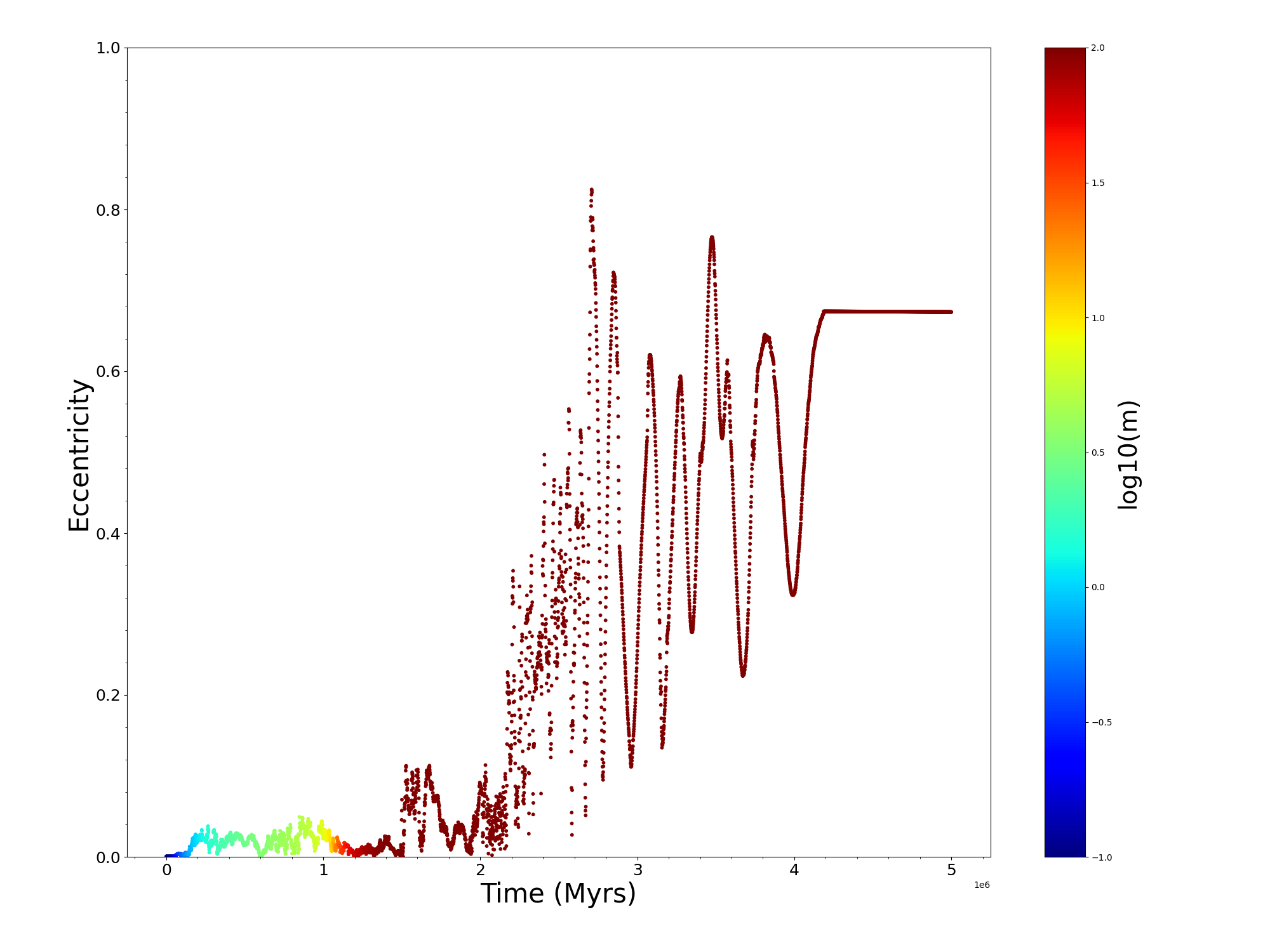}
        \caption{Time correlation of eccentricity with growth as Jupiter grows.}
        \label{fig:evojup_b}
    \end{subfigure}

    \caption['Time Evolution']{Time Evolution of the mass and semi-major axis for the eccentric Jupiter simulation.}
    \label{fig:evojup}
\end{figure}

The cumulative distributions of semi-major axis and eccentricity for the five categories of giant planets are shown in Fig.~\ref{fig:aecdf}. It appears that the Jupiter analogues are approximately evenly spaced in $\log a$, and the same is approximately true for the other types of giant planets, but there are intrinsic differences. The S-J planets are on average the most numerous and closest to the Sun (their median is 10.5~au) and the Jupiter analogues the farthest away (median 16.3 au). The cumulative eccentricity distribution may indicate why this is the case: the Jupiter analogues have, on average, the highest eccentricities, and the I-S planets the lowest. Only 33\% of Jupiter analogues have an eccentricity $e<0.1$ while this fraction is raised to 60\% for the S-J planets and 70\% for Saturn and I-S analogues. It turns out that systems that produce Jupiter analogues are dynamically less stable, leading to an episode of violent scattering. The median semi-major axis for Saturn analogues is 12.9~au, although it reduces to 10.8~au if we restrict ourselves to those Saturn analogues with $e<0.1$. These results indicate that giant planets form throughout the disc with no strongly preferred location and mass configuration (see top panel of Fig.~\ref{fig:orbitdist}), and the outcome in terms of mass and semi-major axis appears random. The eccentricity slope undergoes a notable change at \( e = 0.1 \), the reason for which is unclear, but it seems that systems of multiple giant planets are packed closely enough that eccentricities exceeding 0.1 lead to (near) orbit crossings and a tendency for the systems to become dynamically unstable (see Fig.~\ref{fig:evojup}). The presence of a Jupiter analogue contributes to the prevalence of lower eccentricities.\\

\begin{figure}
  \resizebox{\hsize}{!}{\includegraphics{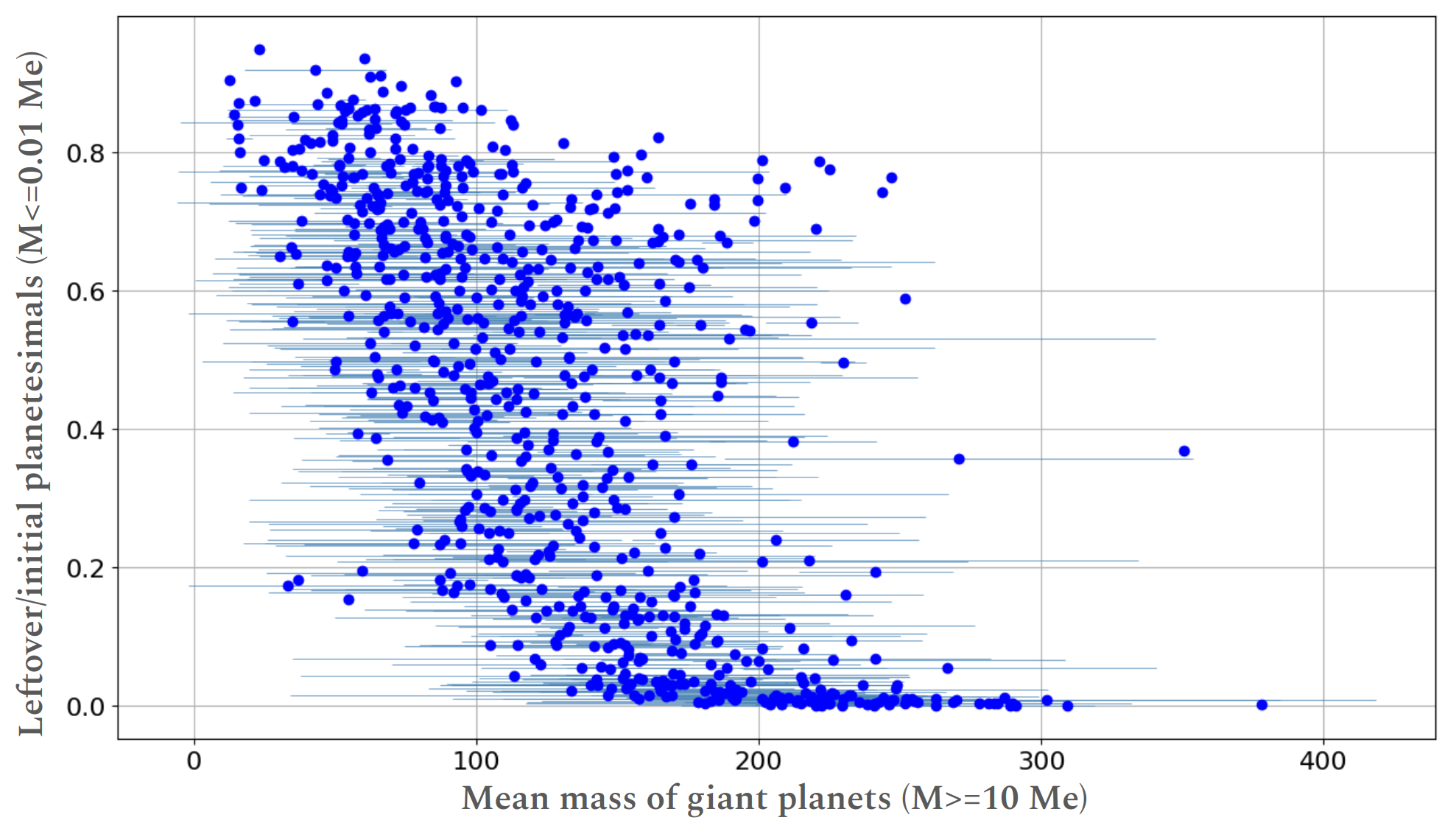}}
   \caption['fracpltms']{Fraction of leftover planetesimals (vertical) versus the mean mass in giant planets ($M > 10 Me$)(horizontal). The normalisation for the vertical axis is done with the initial number of planetesimals for that system. The uncertainty is the standard deviation in giant planet mass. Each dot represents a the mean of giant planet mass from a single simulation. We see an inverse trend for the left-over planetesimals with increasing mass. } 
  \label{fig:fracpltms}
\end{figure}

The higher eccentricity of Jupiter analogues than Saturn analogues is puzzling. We present two potential solutions. The first is that systems with Jupiter analogues have a lower final mass in planetesimals than those without Jupiter analogues, so that the planetesimals cannot damp the eccentricities of the Jupiter analogues. In Figure \ref{fig:evojup} we plot snapshots of the evolution of a simulation that produces a single Jupiter analogue on an eccentric orbit and plot eccentricity vs time for the same. The eccentricity of the giant planets increases due to their mutual scattering, and remains high once the planetesimals are almost completely removed from the system around 3.5 Myr. We back up our claim by showing the final fraction of planetesimals remaining in the simulations versus mean giant planet mass in Fig.~\ref{fig:fracpltms}. We see a rapidly decreasing trend for the fraction of left-over planetesimals, with individual mass less than a lunar mass, with increasing giant planet mass. In other words: the higher the mean mass of giant planets in each simulation, the lower is the fraction of planetesimals remaining. 
A second possibility is hinted at in Fig.~\ref{fig:evojup}: a system of four giant planets becomes dynamically unstable between 3.5 and 4~Myr of evolution. By this time the gas surface density has decreased by at least a factor 33 from the initial value, so we postulate that at that time the eccentricity damping from the gas is insufficient to overcome the mutual perturbations between the giant planets, and the system goes unstable. A dynamical instability has been proposed to explain the fair number of eccentric giant exoplanets that are observed \citep{JT2008}, but there is no reason why this instability has to happen late as opposed to when the gas disc is nearly gone.\\


\begin{figure}
  \resizebox{\hsize}{!}{\includegraphics{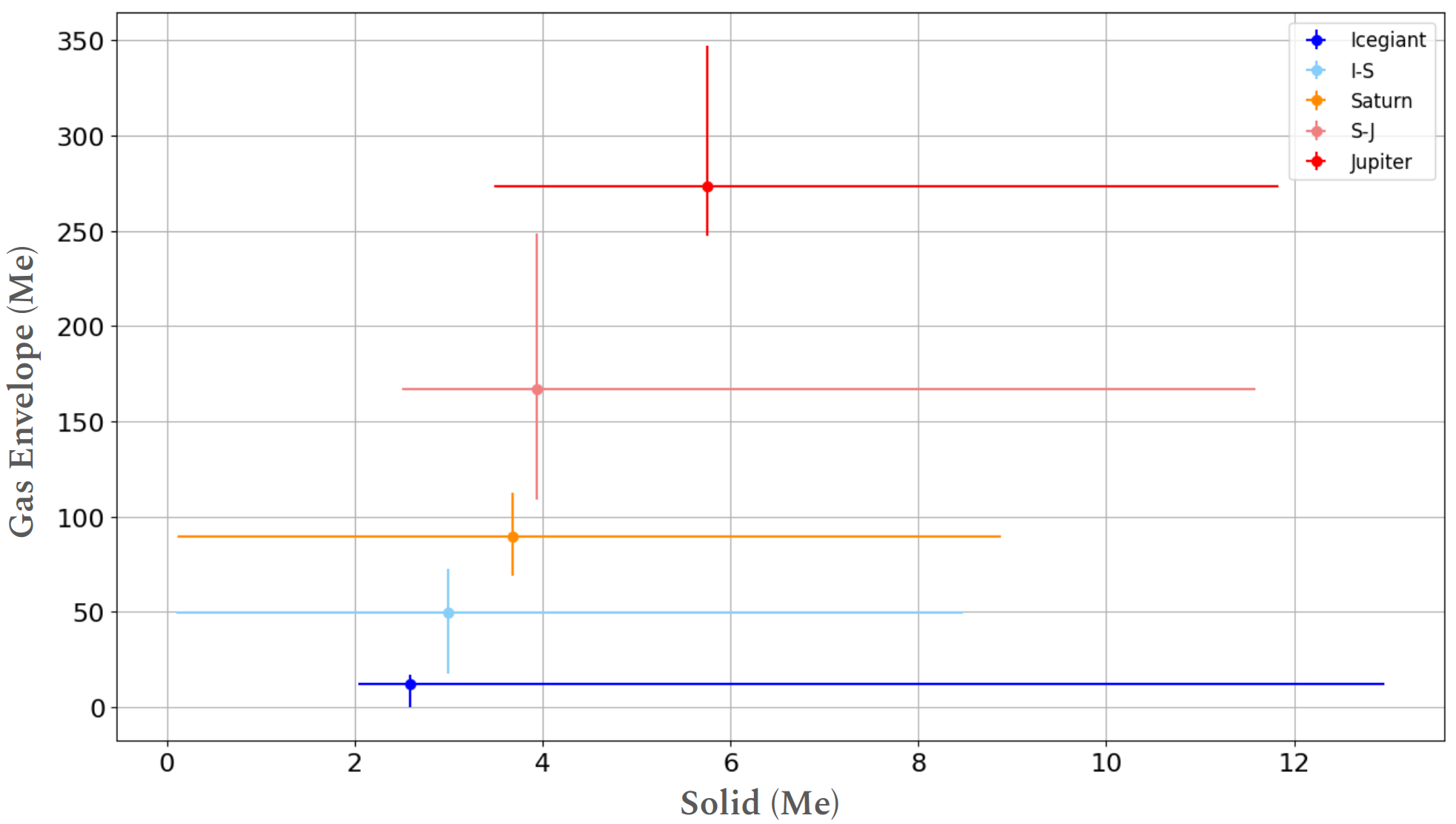}}
   \caption['gtr']{Mass in solids (horizontal) versus mass in gas (vertical) for the five different types of giant planets. The means are plotted. The uncertainties are the min and max values. } 
  \label{fig:gtr}
\end{figure}

In Fig.~\ref{fig:gtr} we plot the mass in solids -- from pebble accretion -- versus the mass in gas for the five types of giant planet that we form. The respective solid to gas fractions of the planetary masses are $0.19 \pm 0.05$ for the ice giants, $0.06 \pm 0.02$ for I-S giants, $0.04 \pm 0.01$ for Saturn analogues, and $0.02 \pm 0.007$ for both S-J and Jupiter analogues. We further computed that planets with a mass of about 4~$M_\oplus$ have a roughly equal mass in core and envelope, consistent with the simulations of \citet{Matsumura2021}. \\

The average number of giant planets with $m_p>10$~$M_\oplus$ at the end of the simulations is $\langle N \rangle = 1.7 \pm 1.24$. The average multiplicity of all multiple planet systems is $\langle N \rangle \approx 2.5$. We produce a few cases of planets in mean-motion resonances, specifically 2:1 ($< 8 \%$), 3:2 and 5:2 ($< 5\%$ each). The above-mentioned percentage value comes from the total cases of resonant pairs divided by total planet pairs formed. However, the occurrence of multiple resonances that form a chain in the same run, even in cases with higher multiplicity, is very rare.

\subsection{Input parameter dependence}
We calculate the success of forming some giant planet analogue through the parameter $\lambda$, which is the number of giant planet analogues per run, and is calculated as

\begin{equation}
    \lambda = \frac{\rm Total\, number\, of\, planets\, formed\, in\, all\, runs}{\rm Total\, number\, of\, runs}
\end{equation}  
This is evaluated for Jupiter, S-J, Saturn, I-S and ice giant analogues. In this section, we will show which values of parameters have the highest success chance in forming Saturn analogues, and also give the best combination of parameters that best reproduces the overall outer solar system. Even though the pebble accretion prescription of \citet{OL2018} is an improvement over that of \citet{Ida2016}, we prefer to show the results of both prescriptions. \\



The fraction of obtaining a Saturn analogue with the Ida pebble accretion prescription is $0.33 \pm 0.26$ ($1\sigma$) while with the Ormel prescription this is $0.32 \pm 0.17$, which are statistically identical. These fractions are  $0.19 \pm 0.16$ when the sticking efficiency $f=0.5$, $0.38 \pm 0.2$ when $f=0.75$ and  $0.35 \pm 0.24$ when $f=1$. These values are also identical within uncertainties. Concerning the initial total mass in planetesimals, then fractions are  $0.21 \pm 0.21$ when $m_{\rm tot}=0.5$~$M_\oplus$,  $0.33 \pm 0.17$ when $m_{\rm tot}=1$~$M_\oplus$ and  $0.34 \pm 0.25$ when $m_{\rm tot}=4$~$M_\oplus$.\\

In contrast, the fraction for forming Jupiter analogues is strongly dependent on the pebble accretion prescription used. With the Ida prescription the fraction is  $0.11 \pm 0.26$, while with Ormel's prescription the fraction for forming Jupiter analogues is $0.02 \pm 0.03$, which is substantially different. Similarly the sticking efficiency plays an important role in the production of Jupiter analogues: when $f=0.5$, we produce no Jupiter analogues because the growth is too slow. When this is increased to $f=0.75$, the fraction increases to  $0.02 \pm 0.06$, but it jumps to $0.17 \pm 0.29$ when $f=1$. Overall the likelihood of obtaining Jupiter analogues and massive gas giants, with mass in between Saturn and Jupiter, decreases with the more detailed accretion efficiency model by \citet{OL2018} and lower sticking efficiency. \\

\begin{figure*}
\begin{centering}
        
    \vspace{1em}
    \begin{subfigure}[b]{\textwidth}
        \centering
        \resizebox{0.8\hsize}{!}{\includegraphics[width=\textwidth]{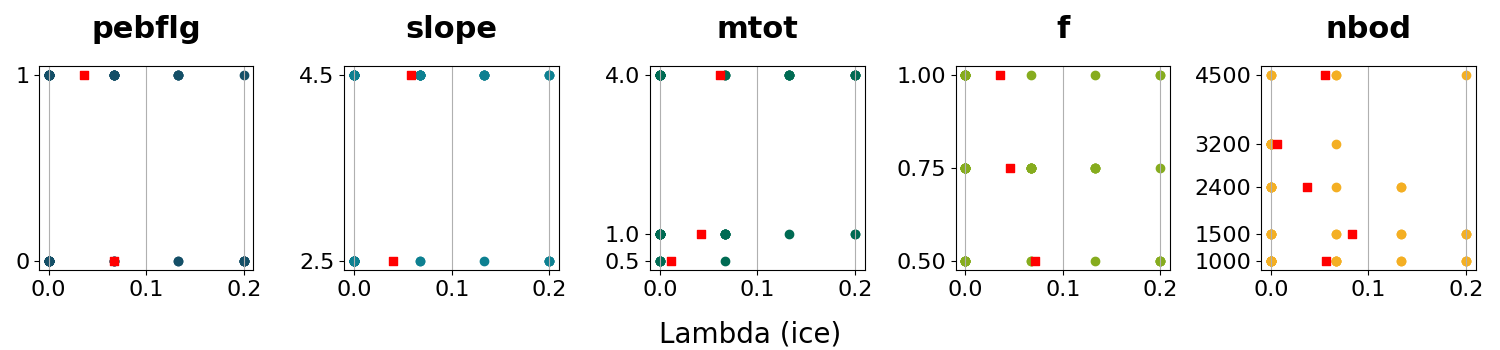}}
    \end{subfigure}
    \vspace{1em}
    \begin{subfigure}[b]{\textwidth}
        \centering
        \resizebox{0.8\hsize}{!}{\includegraphics[width=\textwidth]{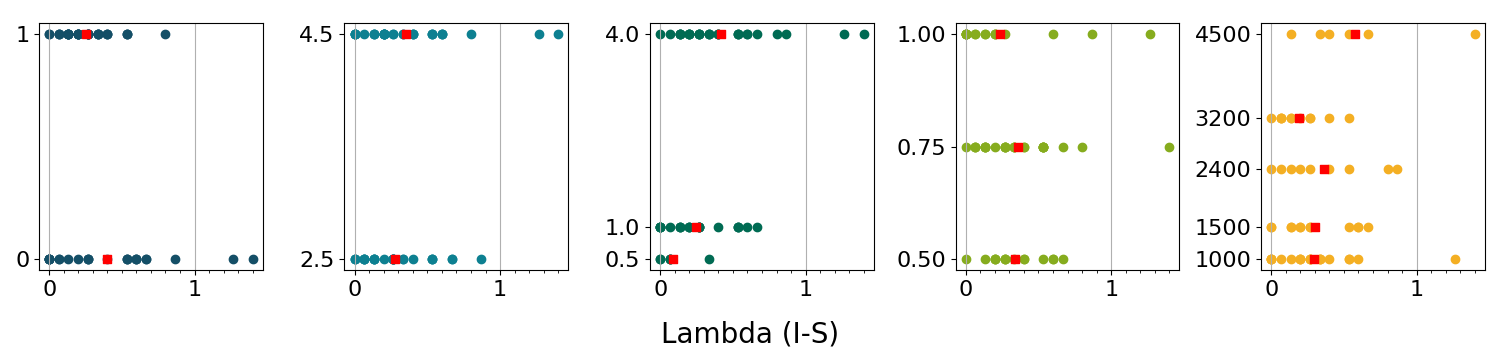}}
    \end{subfigure}
    \vspace{1em}
    \begin{subfigure}[b]{\textwidth}
        \centering
        \resizebox{0.8\hsize}{!}{\includegraphics[width=\textwidth]{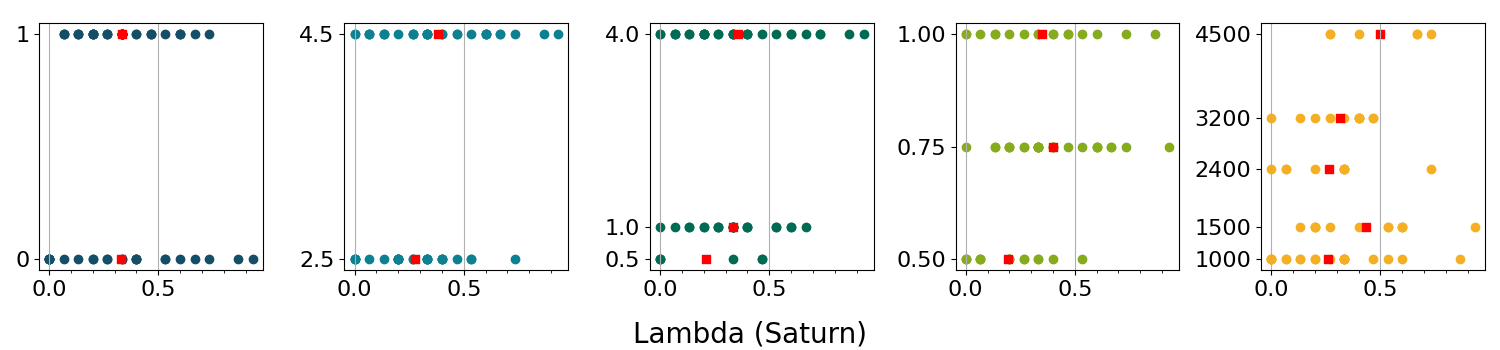}}
    \end{subfigure}
    \vspace{1em}
    \begin{subfigure}[b]{\textwidth}
        \centering
        \resizebox{0.8\hsize}{!}{\includegraphics[width=\textwidth]{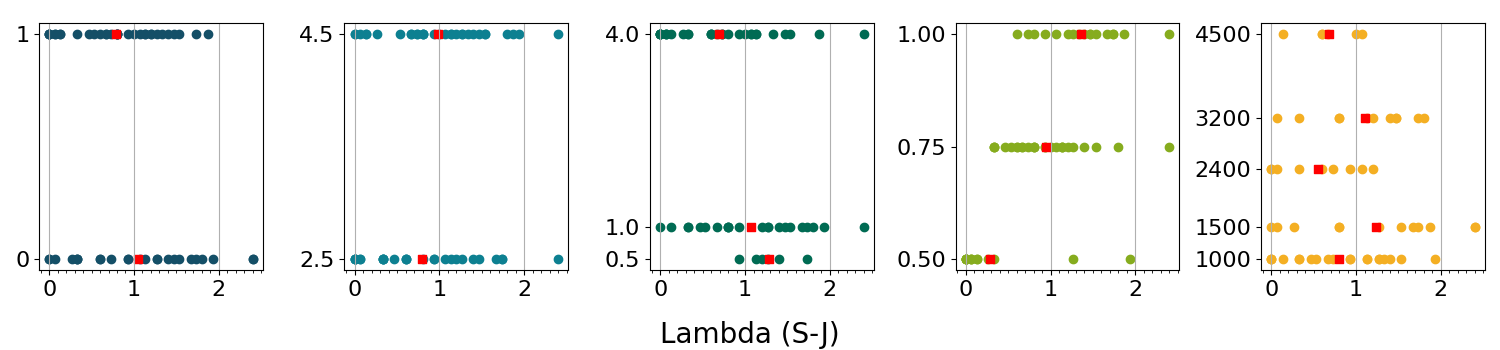}}
    \end{subfigure}
    \vspace{1em}
    \begin{subfigure}[b]{\textwidth}
        \centering
        \resizebox{0.8\hsize}{!}{\includegraphics[width=\textwidth]{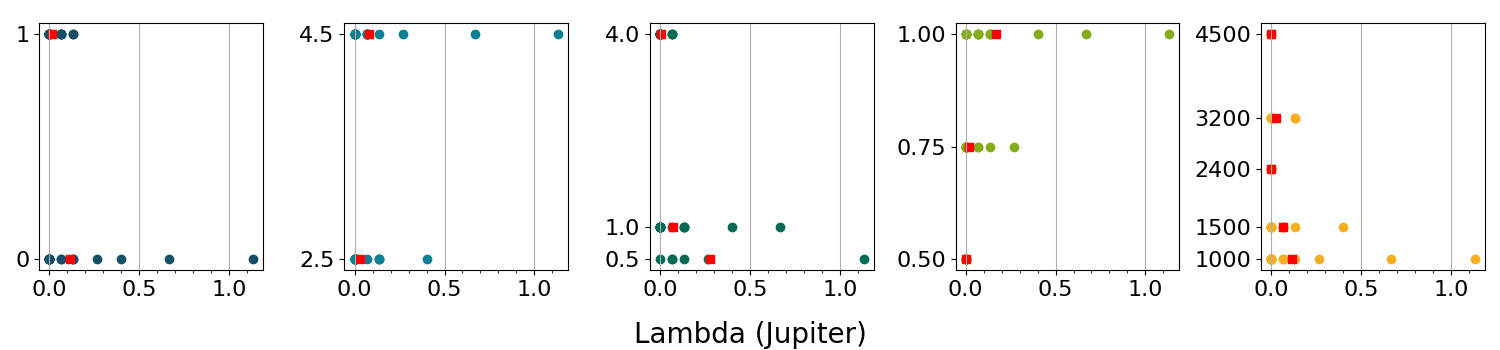}}
    \end{subfigure}
    \vspace{1em}
    \begin{subfigure}[b]{\textwidth}
        \centering
        \resizebox{0.8\hsize}{!}{\includegraphics[width=\textwidth]{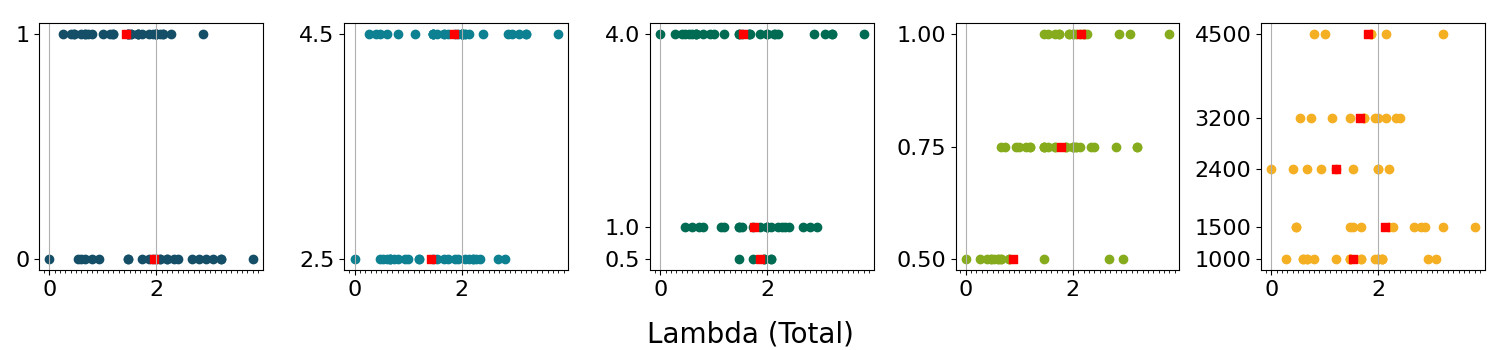}}
    \end{subfigure}

    \caption{Scatter plot for different parameter values vs $\lambda$ for (from top to bottom) Ice giants, I-S, Saturn, S-J, Jupiter mass analogues and all massive bodies. Here for each parameter value, the values of other parameters are varying. The red square indicates the mean of the distribution. There is no single parameter that stands out for any planets formation. Please note that the x-axis limits are varying.}
    \label{fig:lambda}

\end{centering}
\end{figure*}

This dependence of the average number of Saturn analogues (and other mass analogues) on different parameter values can be seen in the scatter plots (Fig. \ref{fig:lambda}). To check whether different values of the same parameter significantly affect the probability of Saturn analogue formation, we ran the K-sample Anderson-Darling test (Anderson \& Darling, 1952) for all possible parameter combinations. This test checks whether several collections of observations can be modelled as coming from a single population, and where the distribution function does not have to be specified. It is more sensitive at the tail ends of the distribution than the Komolgorov-Smirnov test. Given this, we also performed the standard Kolmogorov-Smirnov (KS) test. The results of the two tests are fairly consistent with each other, so only one is reported here. To eliminate strong statistical biases we have only tested those parameters for which we have more than ten sets of simulations. These parameters are: $m_{\rm disc}=1$ and 4~$M_\oplus$, and $N=1000$ and 1500, and all other values of the pebble accretion prescription, $f$, and $\beta$. We found only a few combination of parameters where the $p$-value of the $\lambda$ distribution was below 5\% for Saturn analogues, suggesting that no single parameter has a dominant effect on the efficiency of Saturn analogue formation. In other words, the distribution of outcomes are not statistically different when adopting different input parameters e.g. the outcomes with the Ormel and Ida prescriptions, or the initial slope of distribution. The only combinations that yielded $p<0.05$ are with  $f=0.5$ and $0.75$, and $N=1000$, and 1500. For Jupiter and ice giant analogues there are too few outcomes that have $\lambda>0$, so that the outcome of the Anderson-Darling and K-S tests is not very meaningful. Yet, from Figure \ref{fig:lambda} and looking at S-J analogues, we see that Ida's prescription and $f=1$ tend to form more massive planets, leaning towards Jupiter-mass analogues. Conversely, Ormel's prescription with $f=1$ decreases the formation efficiency of ice giants and I-S analogues, while $m_{\rm tot} = 4$~$M_\oplus$ somewhat favours their formation. A slightly significant trend is seen with the number of planetesimals $N$. With $N = 1000$ and 1500, the number of planets produced increase with their mass. Overall the sticking efficiency $f$ has the most significant effect on the likelihood of forming different types of planet analogues, because in essence it is another parameterisation of the pebble flux.\\


\begin{figure*}
  \begin{centering}  
  \resizebox{\hsize}{!}{\includegraphics{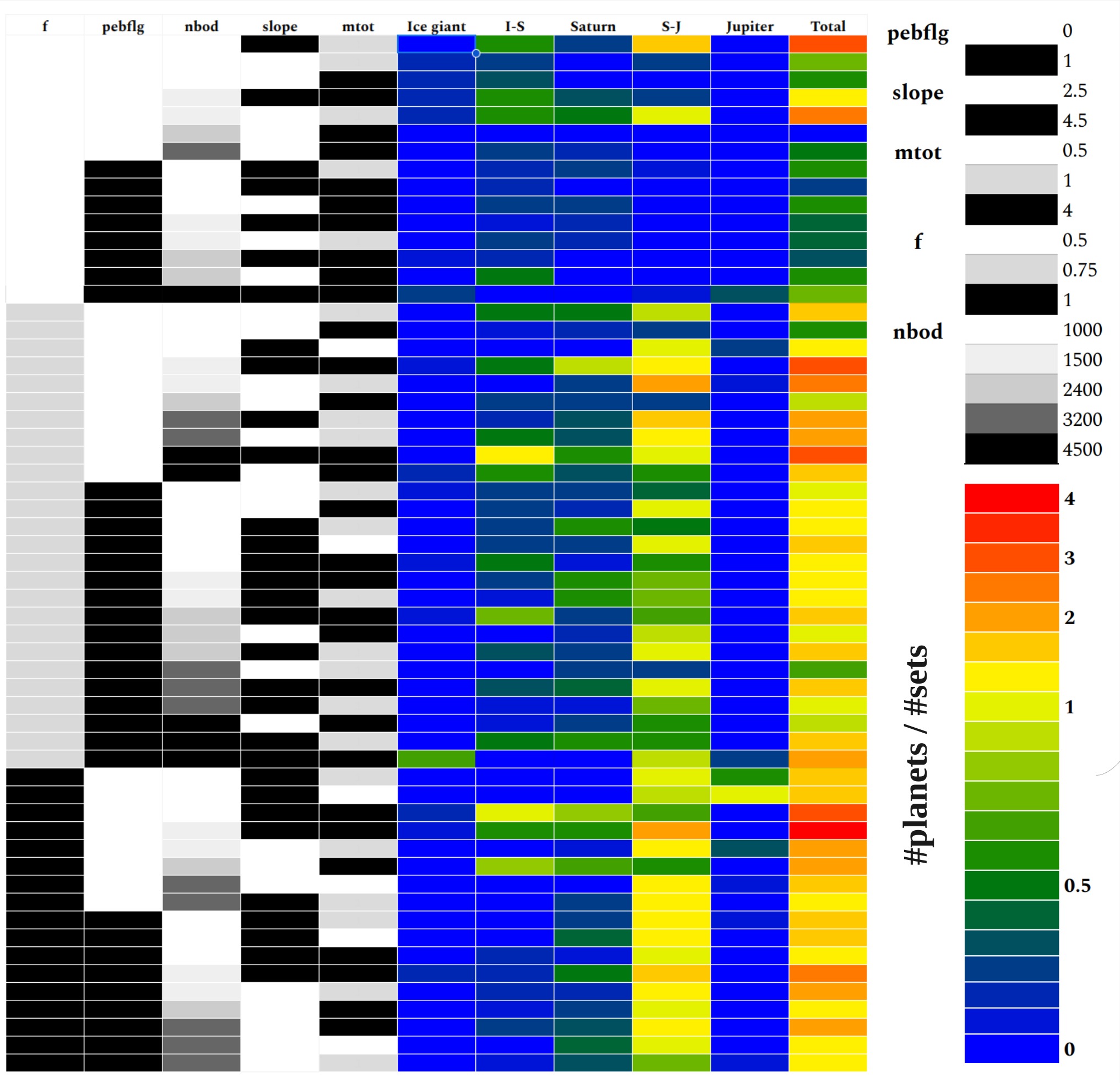}}
  \caption['probcolor']{Summary of all the parameter combinations and their respective $\lambda$ values for each planet. The chart is sorted according to $f$, followed by $\rm peb\_flg$ and $N$. The color bar is deliberately non-uniformly colored to enhance comprehension of the likelihood distribution across the range of values spanning from 0 to 1. }
  \label{fig:probcolor}
  \end{centering}
\end{figure*}

Figure \ref{fig:probcolor} summarises the likelihood of forming different planets for different parameter combination. Although we do not find one single most significant parameter, the chart is sorted according to the most sensitive parameter, $f$, followed by the pebble accretion prescription and the initial number of bodies, $N$. 

\subsection{Gas giant growth time distribution}
\begin{figure}
\resizebox{\hsize}{!}{\includegraphics{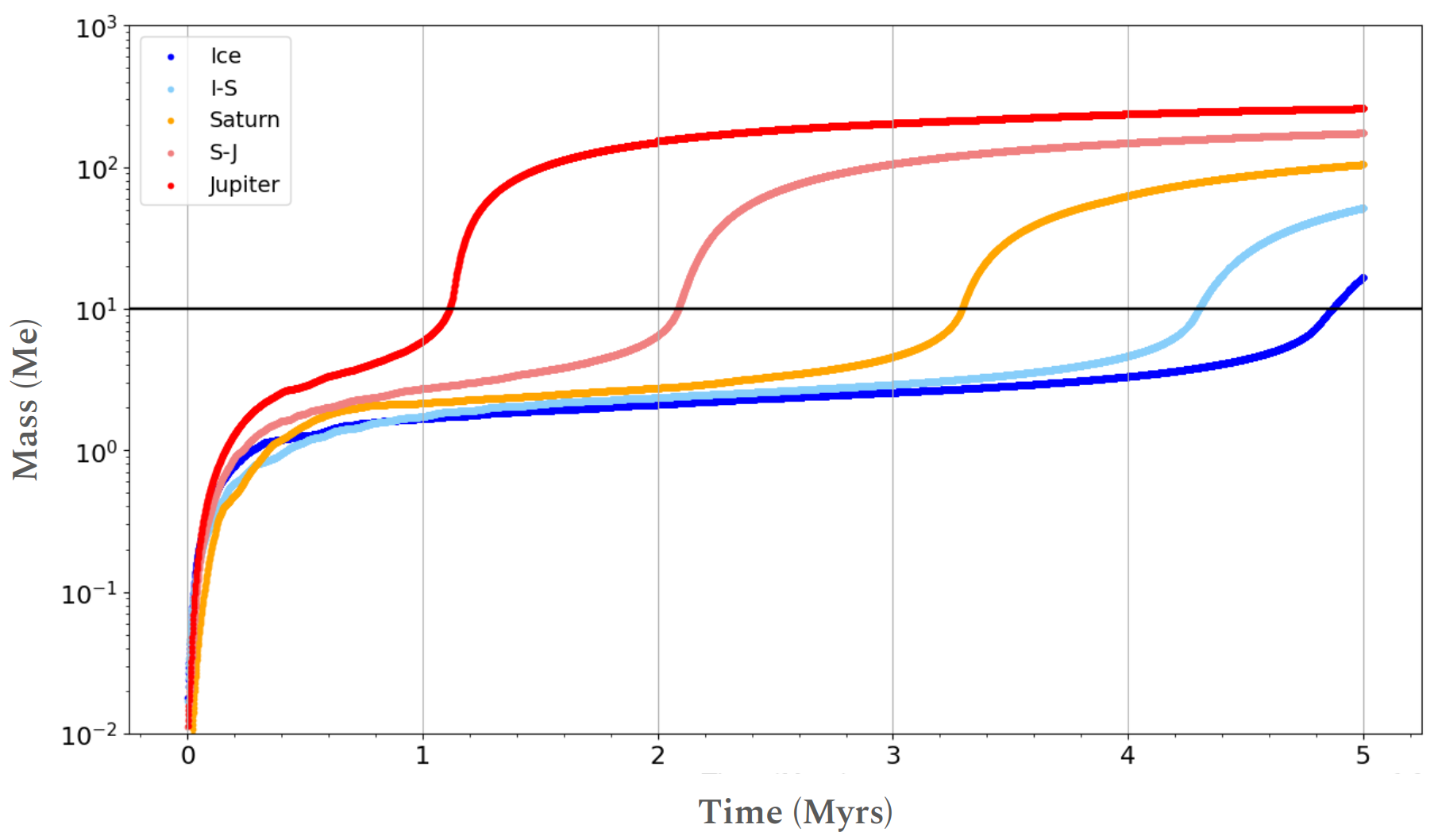}}
\caption['growth']{Tracking growth of all the 5 types of planets over 5 Myr with their typical core formation times. The semi-major axes of each case was close to the mean value for that specific kind of planet. Sequential core formation is discernible.}
\label{fig:growth}
\end{figure}

In this subsection, we analyse the time of formation for all types of gas giants. Fig.\ref{fig:growth} tracks the growth of planets in mass over 5 Myr for a semi-major axis that is close to the mean value for each type of planet. All the planets start from embryos with mass $\leq 0.01 M_\oplus$ which grow to $1 M_\oplus$ within 0.4 Myrs. After this time the growth rate for some bodies slows down considerably. The figure shows that the planets reach 10~$M_\oplus$ cores in a sequential manner: Jupiter analogue cores form by 1 Myr while ice giant analogue cores form the last at around 4.8 Myrs. \\
 
We calculate the time it takes each giant planet to reach 10~$M_\oplus$. There are several reasons for us doing so. First, because it was traditionally considered to be a problem to reach this mass from planetesimal accretion within the disc's lifetime in this region of the Solar System \citep{Pollack1996}. Pebble accretion has clearly alleviated this problem. In addition, the average pebble isolation mass across the giant planet region is about 10~$M_\oplus$ and Fig.~\ref{fig:gtr} shows that 10~$M_\oplus$ is also the approximate upper mass at which gas accretion sets in. The pebble isolation mass is thought to be the mass at which the planet creates wakes in the disc that halt further accretion \citep{Lambrechts2014}, and a successful link has been made between the core formation time of Jupiter and the radiometric Hf-W ages of iron meteorites \citep{Kruijer2017}. Here we build on these results. \\ 

\begin{figure}
\resizebox{\hsize}{!}{\includegraphics{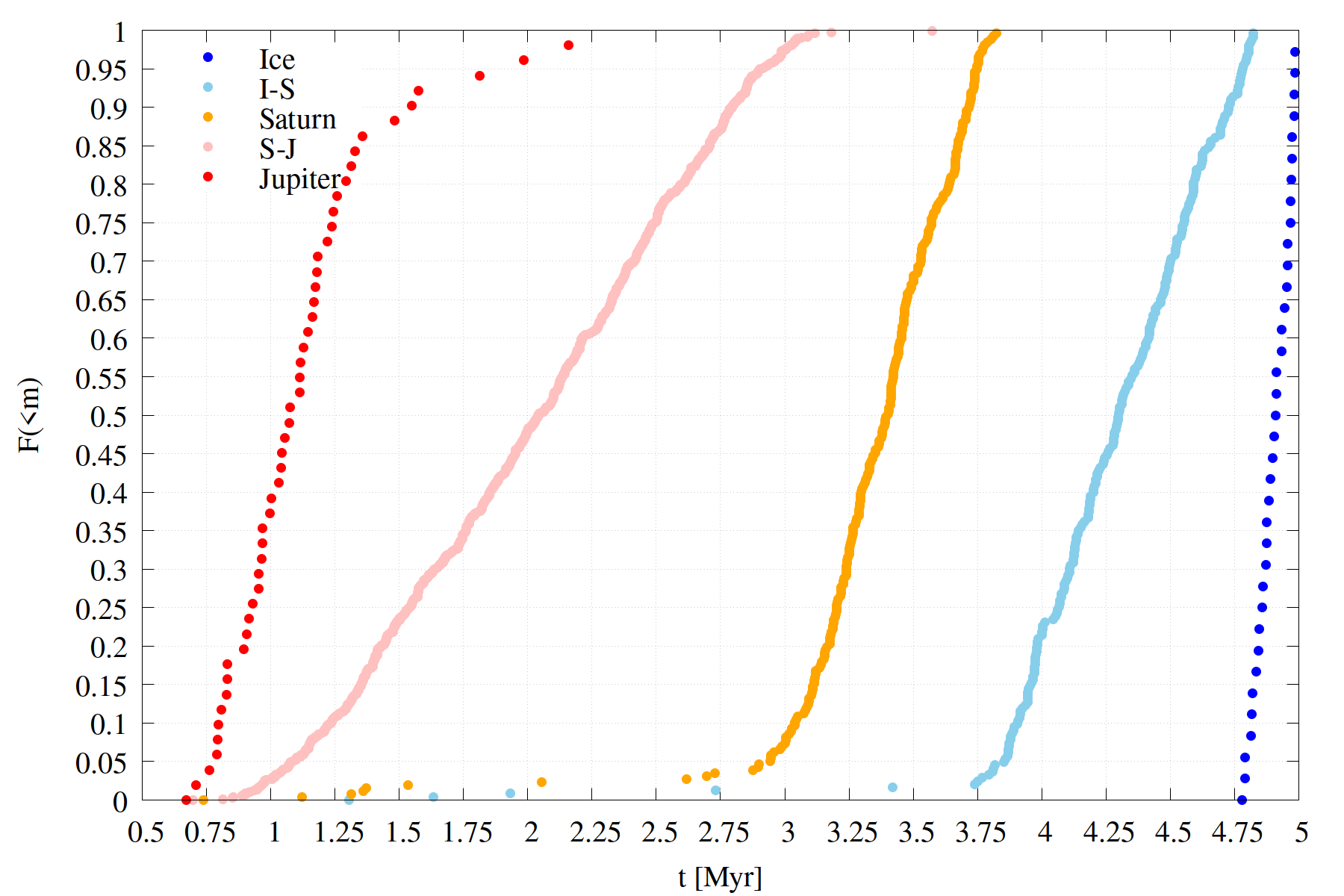}}
\caption['tcdf']{Cumulative distribution for time of accretion to form 10 $M_\oplus$ cores for all five different giant planet analogues. The planets form in distinct time intervals.}
\label{fig:cdftime}
\end{figure}

In Fig.~\ref{fig:cdftime} we show a cumulative distribution of the time it takes for a planetesimal to grow to 10~$M_\oplus$ that will eventually end up as one of the various kind of giant planets. The Jupiter analogues grow the fastest, reaching 10~$M_\oplus$ in $\langle t_{\rm c,J} \rangle= 1.1 \pm 0.3$ Myr. For Saturn analogues this takes considerably longer: $\langle t_{\rm c,S} \rangle= 3.3 \pm 0.4$ Myr, and for the ice giants it takes $\langle t_{\rm c,I} \rangle= 4.9 \pm 0.1$ Myr. These results indicate that the three types of giant planet in the Solar System formed at distinct, non-overlapping times. The two other types of giant planet, not present in the Solar System, have mean core growth times of $\langle t_{\rm c,SJ} \rangle= 2.0 \pm 0.6$ Myr and $\langle t_{\rm c,IS} \rangle= 4.3 \pm 0.4$ Myr. In summary: the cores of the giant planets of the Solar System formed in sequence.\\

\begin{figure}
\resizebox{\hsize}{!}{\includegraphics{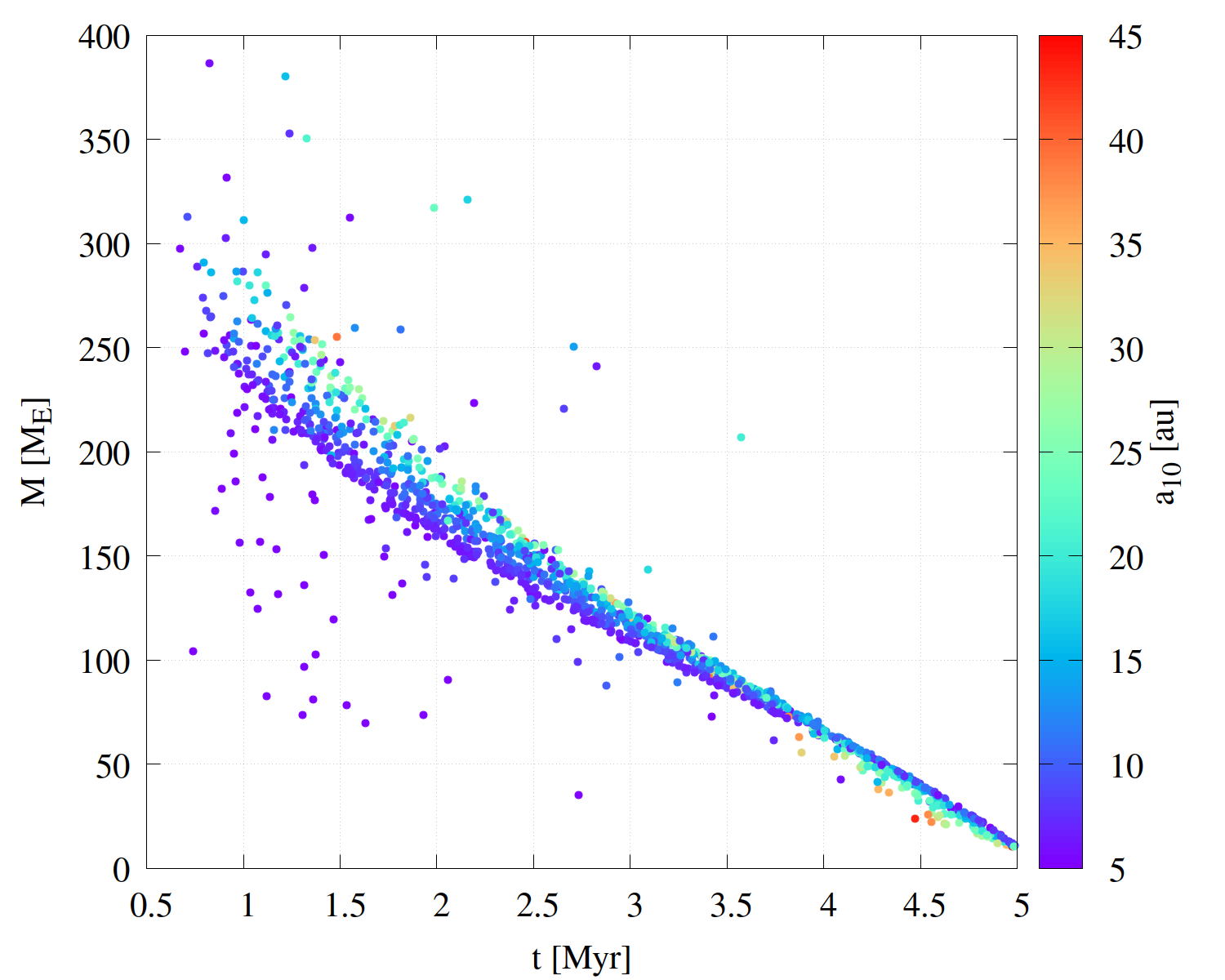}}
\caption['tscatter']{Final mass reached at 5 Myr vs time of accretion to form $10 M_\oplus$. Colours indicates the semi-major axis at which the planet core formed. An inverse relationship is seen.}
\label{fig:formtime}
\end{figure}

The final mass attained by a planet core with a mass of $10 M_\oplus$ as a function of its formation time was also examined, and is illustrated in Fig.~\ref{fig:formtime}. The figure reveals a roughly linear inverse relationship, albeit with large scatter when $t\lesssim 2$~Myr, between the accretion time and the final mass achieved by a 10~$M_\oplus$ core. Faster core formation results in a more massive planet after 5 Myr. \\

\begin{figure}
\resizebox{\hsize}{!}{\includegraphics{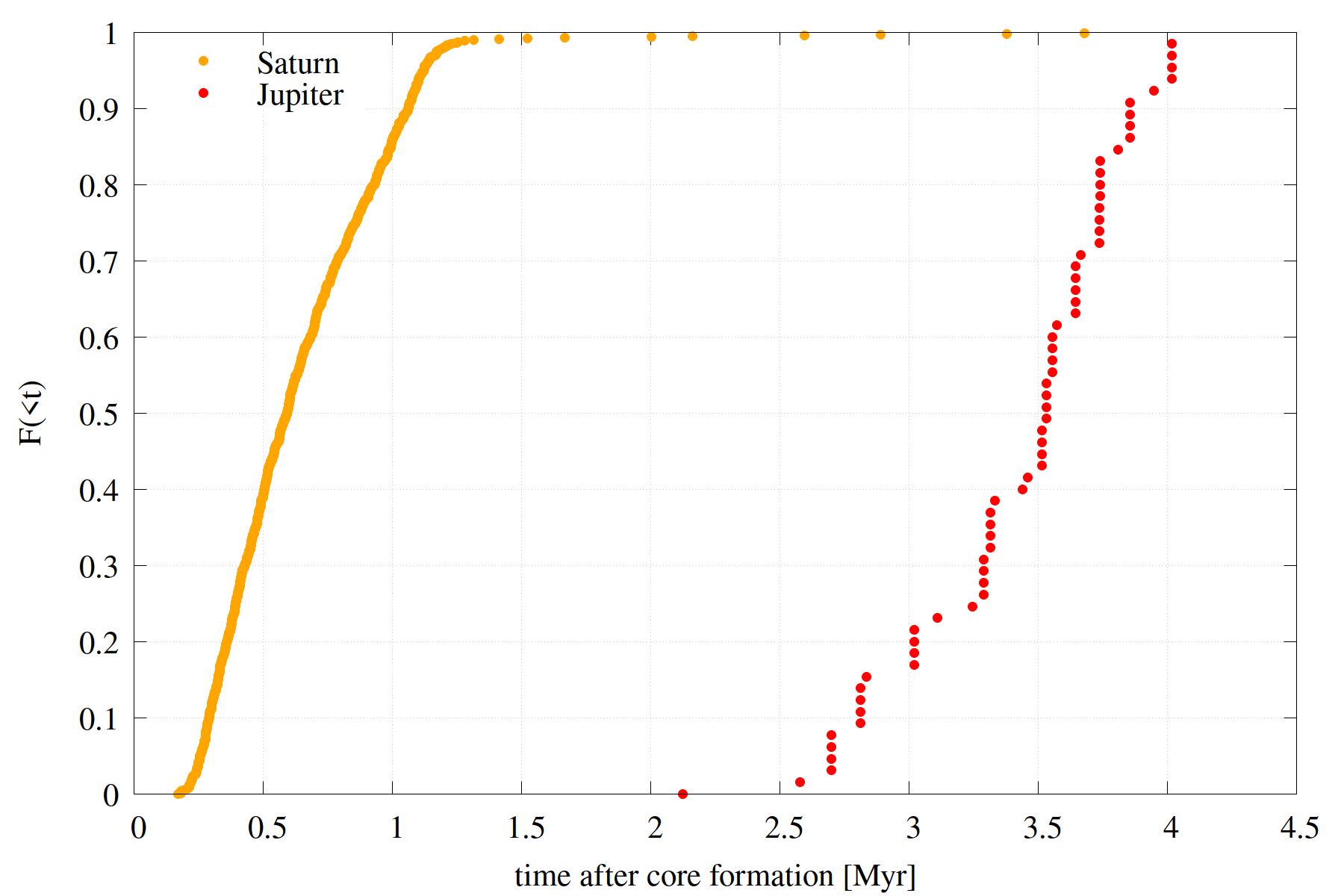}}
\caption['tgas']{Cumulative time distribution after core formation for the Saturn and Jupiter analogues to reach 80\% of their current mass. While Saturn reaches 80\% of it mass after core formation faster than Jupiter, both planets reach their final masses around the same absolute time. }
\label{fig:gastime}
\end{figure}

Last, we investigated how long it takes for the Saturn analogues to reach 75~$M_\oplus$ and the Jupiter analogues to reach 254~$M_\oplus$. This is shown in Fig.~\ref{fig:gastime}, where we plot the time that it takes after core formation for Jupiter and Saturn analogues to reach 80\% of their mass. The Saturn analogues reached 80\% of their final mass much faster than the Jupiter analogues. Indeed, the Saturn analogues reach 75~$M_\oplus$ in $\langle t_{\rm gas,S} \rangle = 0.64 \pm 0.33$~Myr after core formation, while for Jupiter it takes $\langle t_{\rm gas,J} \rangle = 3.37 \pm 0.45$~Myr. Saturn's core forms later, but it finishes growing at more or less the same time as Jupiter. Both planets reach their final masses in 4 to 4.5 Myr, which is about the time the disc is thought to have dispersed \citep{Wang2017}.\\

In summary, our results indicate that the giant planets in the Solar System could have formed sequentially, with the Jupiter analogues forming first, and ice giants only arising at the very end of the simulation, with a formation time very close to 5 Myr, and Saturn analogues are formed from 10~$M_\oplus$ cores after 3 Myr.

\section{Discussion} \label{discuss}
At the end of the simulations we have many Mars- and Earth-mass bodies. These bodies do not exist in the giant-planet region current Solar System so they need to be removed from the simulations at some point in time. Most of these will be ejected once the simulations are carried on for longer, and many of them have already been removed in the more violent simulations. The long term evolution of these simulations will be investigated in a separate publication. \\

We also report having difficulty reproducing the observed mass-semimajor axis relationship of the giant planets. As stated previously, we do not produce all three types of giant planet in the same simulation. This study started as an investigation to form Saturn and was then generalised. As shown, with the parameters we have chosen it is very difficult to form the ice giants: their low gas mass implies growth at the very end of the gas disc's lifetime, suggesting that either some fine-tuning is needed, or that our model does not capture the full possible distribution of outcomes. In future work this will be investigated in more detail.

\subsection{Jupiter analogues with a high semi-major axis}
We observed the formation of Jupiter analogues as far as 16.3 au from the Sun, which is on average farther than the Saturn analogues (see Fig.~\ref{fig:aecdf}). As suggested by \citet{Thommes1999}, giant planets in compact arrangements may undergo a dynamical instability, thereby augmenting their eccentricities. \citet{Thommes2008} showed that  Jupiter and Saturn can swap their orbits, so that it is not necessary to have Jupiter form inside of Saturn. \\

\subsection{Comparison with chronology from the meteorite record}
Here we briefly discuss how our simulations compare with the cosmochemical meteorite record. A more in-depth analysis is done in a separate publication. \\

Our numerical results indicate that the gas giants have almost fully formed by about 4 Myr after the start of the simulations. By itself this is interesting, but ideally the model should be anchored to chronological data from the meteorite record. Is there any such evidence that supports or invalidates our model? \\

From our simulations, the time it takes for the formation of Jupiter's core is comparable to the time taken as inferred from $^{182}$Hf-$^{182}$W systematics and modelled accretion ages of the various meteorite types (i.e., non-carbonaceous and carbonaceous) of iron meteorite parent bodies \citet{Kruijer2017}. In contrast, \citet{Johnson2016} suggested that the growth of Jupiter is reflected in the timing of the impact resetting of the chondrules of the CB (Bencubbin-like) carbonaceous chondrites, which occurred around $\sim$5 Myr after CAIs \citep{Krot2005}. This timing was however revised recently from $^{182}$Hf-$^{182}$W compositions of CH and CB chondrites that indicate that the impact occurred at 3.8 $\pm$0.8 Myr after the formation of Ca-Al-rich Inclusions (CAIs) \citep{Wolfer2023}. The formation of CB, CH, and CR chondrite parent bodies therefore took place up to $\sim1$~Myr earlier than previously proposed based on Pb-Pb chronology of CB chondrules \citep{Krot2005,Bollard2015}. However, the core of our Jupiter analogues formed much earlier, and our Jupiter analogues have reached a mass of approximately 250~$M_\oplus$ after 3.8~Myr. As such, our simulations align better with the suggestions of \citet{Kruijer2017} that the separation of the NC and CC reservoirs, as constrained from accretion ages of iron meteorite parent bodies, may correspond to the formation of Jupiter core. The later 3.8 Myr metal-silicate fractionation age recorded in CB, CH and CR chondrites may be associated with other high-velocity impact events in the outer solar system. \\

For the Jupiter analogues the most important free parameter is the initial mass in planetesimals, probably because there are more high-mass planetesimals in the disc that can rapidly accrete pebbles and enter a runaway growth phase. There is an increased role for the different pebble accretion prescriptions for Jupiter, and the best parameters for producing Jupiter analogues are Ida's pebble accretion prescription, number of planetesimals $N=1500$, and sticking efficiency $f=1$. These parameters tend towards optimisation for the fastest possible growth. For these parameter combinations Jupiter's growth is consistent with that advocated by \citet{Kruijer2017}.
 
\begin{figure}
\resizebox{\hsize}{!}
{\includegraphics{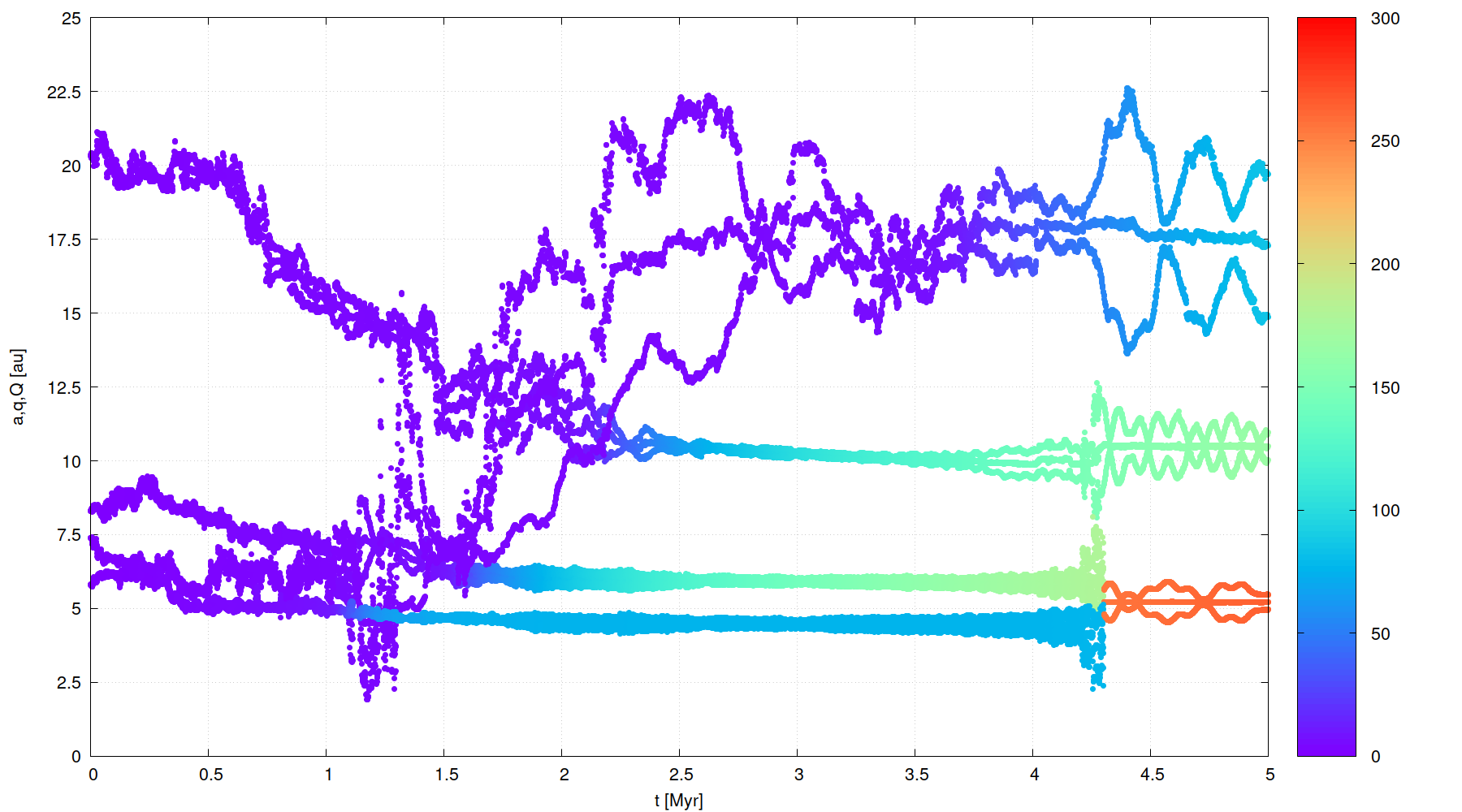}}
\caption['evo']{Evolution of the system displayed in the leftmost middle panel of Fig.~\ref{fig:maic16}. This system yields both a Saturn and a Jupiter-mass planet. The Jupiter analogue is the result of a merger near 4 Myr and 5 au.}
\label{fig:evo}
\end{figure}

\subsection{A few words about composition}
In section 4.2 we reported the solid to gas fractions of the planets that we form in our simulations. We find that the solid masses are generally $<8$~$M_\oplus$, which could be inconsistent with the heavy element enrichment of the gas giants.\\

The actual inner core mass of Jupiter appears to be constrained to at most 7~$M_\oplus$, consistent with our findings, while its total mass in heavy elements can be up to 30~$M_\oplus$ \citep{Miguel2022}. For Saturn the total mass in heavy elements could be up to 18~$M_\oplus$ \citep{Movshovitz2020}. That said, the nitrogen isotopic composition of the gas giants is identical to that of the Sun and thus presumed to be that of the protoplanetary disc from which they accreted \citep{Marty2011,FuriMarty2015}. The D/H ratio of Jupiter is 50\% higher than that of Saturn, with the latter coinciding with estimates of the protosolar nebula \citep{Pierel2017}. A substantial heavy element enrichment of Jupiter's envelope, say $>10\%$ of its mass, through planetesimal accretion with chondritic N isotopes would increase the N isotopic composition of Jupiter beyond its observed range \citep{Marty2011}. Similar arguments hold for Saturn when using its D/H ratio. Furthermore, the D/H ratio of the water vapour in Enceladus' plume is similar to that found for comets, which is higher than that of chondritic meteorites \citep{Waite2009}. This would imply that either Saturn did not accrete a lot of solids, or the atmosphere remains stratified and the D/H measurements are not indicative of the deeper interior. What all of this implies is that formation models may require revision \citep{Alibert2018,Helled2022}.

\subsection{Comparison with previous works}
Here we briefly compare our works with that of previous studies. We focus on those studies that have incorporated pebble accretion only. Much of our methodology is based on the studies by \citet{Matsumura2017,Matsumura2021} so we shall not include those two studies here.\\

One issue that is important is the timing of giant planet core formation relative to disk dissipation. In published planet formation models that consider a smooth disk, e.g. \cite{Bitsch2019,Matsumura2017,Matsumura2021,Lau2024}, the cores of the giant planets generally migrate rapidly towards the Sun resulting in the formation of hot-Jupiters. \citet{Levison2015} avoided this problem by not accounting for planet migration. To form Solar System-like giant planets, the smooth-disk model requires a core formation time that ceases just before disk dissipation so that the strength of planet migration is weak and the total radial amount of migration is limited. However, this formation scenario would contradict the time constraint on the formation of Jupiter from the meteorite record as discussed above. Our results are consistent with other planet formation models that consider disk substructure -- e.g. \citet{Chambers2021} and \citet{Lau2022} -- where a migration trap due to the disk substructure can effectively retain planetary cores from migration. This allows an early formation of Jupiter well before disk dissipation in particular.\\

\citet{Levison2015} studied giant planet with pebble accretion. Unlike us, they added and integrated small particles that were akin to the pebbles. Their major conclusion is that the pebble flux needs to be low enough so that planetesimals gravitationally interact and while they grow the largest ones perturb smaller ones sufficiently so that the latter have higher inclinations and their pebble accretion efficiency is greatly diminished. \citet{Levison2015} begin their simulations with a disc of planetesimals between 4~au and 15~au, and pebbles between 4~au and 30~au, so that our planetesimal disc is more extended. Their simulations lasted for 10~Myr, twice as long as ours. Their initial gas surface density ranges from half to four times higher than ours, and decreases exponentially with an e-folding time of 2~Myr while ours decreases approximately as $t^{-3/2}$. Their disc is also stretched across a greater vertical distance, with their scale height being almost twice as large as ours, while they set the Stokes number to a fiducial value and ours changed with time and distance. Their study included eccentricity and inclination damping, but no migration. Last, their gas accretion parameters are $\log b=9$ and $c=3$, but they limit the gas accretion rate to the Bondi value. Their temporal pebble flux through the disc is based on dust to pebble conversion and follows a different profile from ours. As such, even though we based some of our setup on theirs, their gas disc parameters, gas accretion parameters and temporal pebble flux are very different. Nevertheless, their number of gas giants is similar to ours, but they obtain more ice giants than us. This could be an artefact of the different gas accretion rate and maximum planetesimal diameter.\\

\citet{Chambers2021} studied giant planet formation with pebbles and in a disc with multiple pressure bumps. Using a numerical model that includes migration, gap opening, frictional damping and turbulent excitation of the inclination and eccentricity. The gas disc has eight logarithmically spaced bumps. The simulations place Ceres-mass embryos at the pressure maxima, comparable to the largest objects formed with the streaming instability. \citet{Chambers2021} concludes that gas giant planets can grow extremely rapidly, within 1~Myr, at each pressure bump, even out to 60~au, but are sensitive to initial embryo mass and disc turbulence. Furthermore, massive planets open deep gaps and the maximum gas giant mass is comparable to that of Jupiter. Even though we do not include gap opening in our simulations, the formation timescale of our Jupiter analogues and our maximum planet mass being around 400~$M_\oplus$ are consistent with this study. It is not clear how many Saturn analogues \citet{Chambers2021} produces.\\

\citet{Lau2024} performed a comparable study before this one; it relied on the same software and pebble accretion prescriptions, but differed in the initial planetesimal mass distribution as well as the evolution of the stellar accretion rate and thus the corresponding temporal pebble mass flux. How the outcome of the simulations depends on the initial planetesimal mass is a topic of debate. They also used different values for the pebble isolation mass and gas accretion parameters, and considered sequential gas accretion, which we did not consider. Similarly to \citet{Levison2015}, \citet{Lau2024} concluded that stirring of the planetesimals and embryos is important in limiting their growth from pebble accretion, and that when migration is not considered they produce 1-2 gas giants. If migration is included then planetary cores migrate to the inner disc. This is one of the reasons we investigated the effect of truncating the disc -- and therefore migration -- at 5 au. In agreement with \citet{Lau2024}, we conclude that the pebble accretion prescription of \citet{OL2018} is less efficient at producing giant planets than that of \citet{Ida2016}.

\section{Summary and conclusions} \label{summary}
This study investigates the formation and evolution of giant planets through N-body simulations using the parallel version of SyMBA \citep{LL2023}. The simulations include damping forces from the gas disc, pebble and gas-envelope accretion, and various parameter values to replicate realistic conditions. A total of 840 simulations were conducted with different parameter combinations, initially focused on understanding Saturn's formation, but ultimately to understand giant planet formation as a whole. The simulations ran for 5 million years, tracking the dynamics of self-gravitating planetesimals. The computational efficiency of the simulations, orbital distribution, core formation time and parameter combinations were systematically tested and analyzed. Key parameters studied include sticking efficiency, pebble accretion prescription, initial total planetesimal mass, size-frequency slope, and the initial number of planetesimals. \\

At the end of 5 Myr, while we observed a significant diversity in the orbital distribution of giant planets, we saw no correlation between semi-major axis, eccentricity and mass of the planet. Apart from Jupiter analogues, most giant planets exhibit low eccentricities ($e < 0.1$), with a small fraction displaying eccentricities exceeding 0.5, indicative of dynamically unstable systems. The cumulative distributions of semi-major axis and eccentricity highlight differences among planet types; however, there is no preferred location or mass configuration for the giant planet formation. Jupiter analogues are on average much more eccentric, possibly due to dynamical instabilities that leave too little mass in planetesimals to damp the eccentricities down.\\

The likelihood of forming a type of planet ($\lambda$) analogues varied based on parameter combinations. Statistical tests suggest no single parameter significantly dominates formation efficiency, indicating a complex interplay among parameters. Nonetheless, the sticking efficiency appears to have the most significant impact on the formation of different types of planet analogues, but not at a 2$\sigma$ level.  \\

Finally, we analyzed the mean formation times of the cores of the giant planet analogues. Jupiter analogues exhibit the shortest growth time (to grow into 10 $M_{\oplus}$), with a mean of $\langle t_{\rm c,J} \rangle= 1.1 \pm 0.3$ Myr, followed by Saturn analogues $\langle t_{\rm c,S} \rangle= 3.3 \pm 0.4$ Myr and ice giants $\langle t_{\rm c,I} \rangle= 4.9 \pm 0.1$ Myr, indicating sequential formation. Additionally, we observe that the final mass attained by a 10 $M_{\oplus}$ core is inversely proportional to its formation time, reinforcing the notion of sequential formation in the solar system. Furthermore, we explore the time taken for Saturn analogues to reach 75 $M_{\oplus}$ and Jupiter analogues to reach 254 $M_{\oplus}$, finding that Saturn analogues reach 80 \% of their final mass earlier than Jupiter analogues, though both reach their final masses in 4 to 4.5 Myr, aligning with the estimated dispersal time of the disc.\\

Our findings provide valuable insights into the timing and mechanisms of giant planet formation, offering implications for our understanding of the early solar system dynamics as well as shortcomings and possibilities for improvement. In the future, we will focus on achieving the formation of Jupiter and Saturn with their current orbital configurations. We aim to explore the intricate substructures of protoplanetary discs, such as rings, in order to better understand their influence on the formation and positioning of these gas giants in the solar system.

\section*{Acknowledgements}
T.C.H.L. acknowledges funding from the Deutsche Forschungsgemeinschaft (DFG, German Research Foundation) under grant 325594231. Man Hoi Lee was supported in part by Hong Kong RGC grant 17306720. We thank Sourav Chatterjee, Bernard Marty and two anonymous reviewers for useful and constructive feedback.
\bibliographystyle{elsarticle-harv} 
\bibliography{example}




\end{document}